\documentclass[11pt]{article}

\usepackage{amsmath,latexsym,amssymb,slashed,mathrsfs,color}
\usepackage[dvips]{graphicx}

\newcommand{\f}[2]{\frac{ #1}{ #2}}
\newcommand{\tf}[2]{{\textstyle\frac{ #1}{ #2}}}

\newcommand{\la}{\langle}
\newcommand{\ra}{\rangle}
\newcommand{\lla}{\la\!\la}
\newcommand{\rra}{\ra\!\ra}
\newcommand{\lella}{\left\la\!\!\left\la}
\newcommand{\rirra}{\right\ra\!\!\right\ra}
\newcommand{\tr}{{\rm tr}\,}

\newcommand{\C}{{\cal C}}
\newcommand{\W}{{\cal W}}

\newcommand{\Op}{{\cal O}}

\newcommand{\slap}{{\slashed p}}
\newcommand{\T}{{\rm T}}
\newcommand{\is}{i_s}
\newcommand{\isp}{i_{s'}'}
\newcommand{\para}{\parallel}

\makeatletter
\@addtoreset{equation}{section}
\makeatother

\begin{document}

\renewcommand{\thefootnote}{\fnsymbol{footnote}}

\vfil
\noindent \hfill Revised version\\
\noindent \mbox{} \hfill November 2015

\vspace*{1.0cm}
\centerline{\bf \LARGE Remarks on the static dipole-dipole potential}
\vskip 0.1cm
\centerline{\bf \LARGE  at large distances} 
\vskip 0.5cm

\centerline{\Large Matteo Giordano$^{\rm a,}$\footnote{e-mail:
    giordano@atomki.mta.hu} and Enrico Meggiolaro$^{\rm
    b,}$\footnote{e-mail: enrico.meggiolaro@unipi.it}}
\vskip 0.3cm

\centerline{\it $^{\rm a}$Institute for Nuclear Research of the
  Hungarian Academy of Sciences (ATOMKI)}
\centerline{\it  Bem t\'er 18/c, H--4026 Debrecen, Hungary}
\vskip 0.1cm
\centerline{\it $^{\rm b}$ Dipartimento di Fisica, Universit\`a di
  Pisa, and INFN, Sezione di Pisa}
\centerline{\it Largo Pontecorvo 3, I--56127 Pisa, Italy}
\vskip 0.75cm

\renewcommand{\thefootnote}{\arabic{footnote}}
\setcounter{footnote}{0}

\begin{abstract}
  We determine the large-distance behaviour of the static
  dipole-dipole potential for a wide class of gauge theories on
  nonperturbative grounds, exploiting only general properties of the
  theory. In the case of QCD, we recover the known results in the
  regime of small dipole sizes, and discuss recent nonperturbative
  calculations. Moreover, we discuss the case of pure-gauge theories,
  and compare our prediction with the available lattice results.
\end{abstract}

\section{Introduction}
\label{sec:intro}

The potential between two static colourless dipoles is the simplest
example of interaction between colour-neutral objects that can be
studied in the framework of non-Abelian gauge field theories. 
The main physical application of this quantity is in the study of the
interaction between quarkonia, i.e., mesons made of heavy quarks,
which can be treated as static colourless dipoles in a first
approximation. 
From the theoretical point of view, the study of the static
dipole-dipole potential poses a nontrivial challenge. Indeed, as one
is typically interested in its large-distance behaviour to describe
the interaction between quarkonia, the interesting properties of the 
potential are mainly affected by the nonperturbative behaviour of the
underlying theory, namely Quantum Chromodynamics (QCD). Another
complication stems from the fact that the mathematical objects
relevant to the theoretical study of the dipole-dipole potential are
nonlocal operators, namely Wilson loops. 

Calculations of the static dipole-dipole potential
available in the literature~\cite{Pot1,Pot2,Pot3,Pot4,FK} deal with
the regime of small dipole sizes in $SU(N_c)$ gauge theories, mostly using
perturbative techniques. Even in this somewhat simpler framework,
the determination of the static potential requires a careful treatment
of colour interactions within the dipoles, and of the nonlocality of
the Wilson loop, in order to avoid the apparent divergence of the
potential. This requires a partial resummation of the perturbative
series~\cite{Pot1,Pot2}, or equivalently a representation of the
static dipoles in terms of a series of local operators, in the spirit
of the Operator Product Expansion~\cite{Pot3,Pot4,FK}.
For short inter-dipole distances $b$, larger than the dipole size $r$ but
smaller than the typical hadronic scale, $r\ll b \lesssim 1~{\rm fm}$,
one can reliably apply perturbation theory to obtain an estimate of
the potential, which behaves as $V_{dd}\sim
1/b^7$~\cite{Pot1,Pot2,Pot3,Pot4,FK}. At large distances, 
instead, one has to supplement the perturbative description of the
small dipoles with nonpertubative techniques, like the chiral
Lagrangians used in Ref.~\cite{FK}. The leading behaviour for $ b \gg
1~{\rm fm}$ is related to the two-pion threshold, and was found to be
of the form $V_{dd}\sim e^{-2 m_\pi b}/b^{\f{5}{2}}$~\cite{Pot4,FK}.

In this paper we want to study the static dipole-dipole potential in a
purely nonperturbative setting, starting from the definition in terms
of a certain Wilson-loop correlation function, and using only general
properties of the theory, namely its symmetries and its spectrum, to
derive the asymptotic large-distance behaviour. The basic idea is
to insert a complete set of states between the Wilson loops in the
relevant correlation function, and relate the large-distance behaviour
of the potential to the spectrum of the theory. 

There are several motivations behind this work. First of all, the
fully general results for the dipole-dipole potential derived in this
paper provide nontrivial benchmarks for approximate nonperturbative
approaches to QCD, like the anti-de Sitter/QCD (AdS/QCD)
correspondence or the Instanton Liquid Model (ILM), 
and to gauge theories in general. In particular, we confirm the
previous calculations of Refs.~\cite{Pot4,FK}, and provide 
a fully nonperturbative definition of the various numerical factors
entering $V_{dd}$. We also compare our results to the recent
determinations of Refs.~\cite{LZ1,LZ2}, based on AdS/QCD and on the
ILM, respectively.
Moreover, since our results apply to a generic gauge theory (with mass
gap), it is possible to obtain information on the interaction of
colour-neutral states in various theoretically interesting limits,
like the isospin limit, or the {\it quenched} limit, and to establish how
sensitive it is to these ``deformations'' of QCD.

The plan of the paper is the following. After setting the notation in
Section \ref{sec:not}, in Section \ref{sec:wlc} we express the static
dipole-dipole potential in terms of a sum over a complete set of
states. In Section \ref{sec:largeb} we study the behaviour of the
potential at asymptotically large distances, focussing in particular
on pure $SU(N_c)$ gauge theory, and on gauge theories with light
fermions (which include QCD). Finally, in Section 
\ref{sec:concl} 
we draw our conclusions. Most of the technical details are reported in
the Appendices 
\ref{app:partitions}, \ref{app:expo}, \ref{app:largeT} and
\ref{sec:spzero}.

\section{Notation}
\label{sec:not}

In this Section we briefly summarise the important points concerning
Wilson loops, and concerning the sum over a complete set of states,
mainly to set the notation.

\subsection{Wilson-loop operators}
\label{subsec:wlo}

In the functional-integral formalism, the Minkowskian Wilson loop
${\W}_M[\C]$ is defined as follows
\begin{equation}
  \label{eq:wlE_FI_M}
  {\W}_M[\C] = \f{1}{N_c} \tr {\rm P} \exp\left\{
-ig\oint_{\C} A_{\mu}(X) d X^{\mu} \right\} \,,
\end{equation}
for a general path $\C$, where ${\rm P}$ denotes
path-ordering,\footnote{Larger path-times appear on the {\it left}.} and
$A_{\mu}$ are (Minkowskian, Hermitian) non-Abelian gauge fields,
taking values in the $N_c$-dimensional defining representation of
the algebra of the gauge group. The case we have in mind is that of
gauge group $SU(N_c)$, but our formalism extends immediately to any
subgroup of the unitary groups. In the operator formalism, the
Minkowskian Wilson-loop operator reads~\cite{BW,CT}   
\begin{equation}
  \label{eq:wlE_op_M}
  \hat{\W}_M[\C] = \f{1}{N_c} \tr {\rm TP} \exp\left\{
-ig\oint_{\C} \hat A_{\mu}(X) d X^{\mu} \right\} \,,
\end{equation}
where ${\rm T}$ denotes time-ordering of the (Hermitian) non-Abelian
gauge-field operators $\hat{A}_\mu(X) = e^{i\hat{H}X^0} \hat{A}_\mu
(0,\vec{X}) e^{-i\hat{H}X^0}$, where $\hat{H}$ is the Hamiltonian operator. 
In this paper we will be concerned only with
rectangular paths. In Minkowski space, we will denote by 
$\C_M(z_M,R_{M},T)$ the paths running along the contour of the rectangles
${\cal R}_M(\sigma,\tau)$, 
\begin{equation}
  \label{eq:WL_path_M}
  {\cal R}_M(\sigma,\tau) =
 z_M + R_M \sigma + T u_M \tau\,, \qquad
\sigma,\tau\in[-\tf{1}{2},\tf{1}{2}]\,,
\end{equation}
where 
\begin{equation}
  \label{eq:WL_path3_M}
u_M=(0,1,\vec 0_\perp)\,, \qquad  R_M = (r_\para,0,\vec r_\perp)\,, \qquad 
z_M = (b_\para,0,\vec b_\perp)\,.
\end{equation}
Notice that here $T$ does not correspond to the time-extension of the
loop, which is $|r_\para|$ instead. 
For the corresponding Wilson loops (at $z_M=0$) we will use the
following notation, 
\begin{equation}
  \label{eq:WL_notation_M}
  {\W}_M^{\,(T)}(r_\para,\vec r_\perp) =     {\W}_M[\C_M(0,R_{M},T)]\,, \qquad
  \hat{\W}_M^{\,(T)}(r_\para,\vec r_\perp) =     \hat{\W}_M[\C_M(0,R_{M},T)]\,.
\end{equation}
The Euclidean Wilson loop for a general Euclidean path $\C$,
denoted by ${\W}_E[\C]$ in the functional-integral formalism, and by
$\hat{\W}_E[\C]$ in the operator formalism, is defined exactly as in
Eqs.~\eqref{eq:wlE_FI_M} and \eqref{eq:wlE_op_M}, except that the fields
and the scalar product are now Euclidean, and time-ordering is with
respect to Euclidean ``time'', which is here the fourth Euclidean
coordinate. Explicitly,
\begin{equation}
  \label{eq:wlE_FI}
  {\W}_E[\C] = \f{1}{N_c} \tr {\rm P} \exp\left\{
-ig\oint_{\C} A_{E\mu}(X_E) d X_{E\mu} \right\} \,,
\end{equation}
in the functional-integral formalism, and
\begin{equation}
  \label{eq:wlE_op}
  \hat{\W}_E[\C] = \f{1}{N_c} \tr {\rm TP} \exp\left\{
-ig\oint_{\C} \hat A_{E\mu}(X_E) d X_{E\mu} \right\} \,,
\end{equation}
in the operator formalism, where $\hat{A}_{E4}(X_E) \equiv
e^{\hat{H}X_{E4}} (-i)\hat{A}_0(0,\vec{X}_E) 
e^{-\hat{H}X_{E4}}$ and $\hat{A}_{Ei}(X_E) \equiv
e^{\hat{H}X_{E4}} \hat{A}_i(0,\vec{X}_E) e^{-\hat{H}X_{E4}}$,
$i=1,2,3$ 
(these operator relations must be understood in the ``weak'' sense,
i.e., they hold for matrix elements of the operators).
The Euclidean rectangular paths analogous to those defined in
Eq.~\eqref{eq:WL_path_M} will be denoted by $\C_E(z_E,R_{E},T)$, and
run along the contour of the rectangles ${\cal R}_E(\sigma,\tau)$ in
Euclidean space, 
\begin{equation}
  \label{eq:WL_path_b}
{\cal R}_E(\sigma,\tau) = z_E + R_{E} \sigma + T u_E \tau\,, \qquad
\sigma,\tau\in[-\tf{1}{2},\tf{1}{2}]\,,
\end{equation}
where 
\begin{equation}
  \label{eq:WL_path3}
 u_E=(1,\vec
  0_\perp,0)\,,\qquad   R_E = (0,\vec r_\perp,r_\para)=(0,\vec r)\,,
  \qquad z_E=(0,\vec b_\perp,b_\para) = (0,\vec b) \,.
\end{equation}
For the corresponding Wilson loops (at $z_E=0$) we will use the
following notation, 
\begin{equation}
  \label{eq:WL_notation}
  {\W}_E^{\,(T)}(r_\para,\vec r_\perp) =     {\W}_E[\C_E(0,R_{E},T)]\,, \qquad
  \hat{\W}_E^{\,(T)}(r_\para,\vec r_\perp) =     \hat{\W}_E[\C_E(0,R_{E},T)].
\end{equation}
The Euclidean and Minkowskian Wilson loops $\hat{\W}_E^{\,(T)}$ and
$\hat{\W}_M^{\,(T)}$ can be formally related by analytic continuation.
Indeed, the gauge fields in the
Euclidean and Minkowskian Wilson loop appear only in the combinations
$\hat{A}_{E\mu}(X_{E})  dX_{E\mu}$ and
$\hat{A}_{\mu}(X)  dX^\mu$, respectively,
which are formally related as follows:
\begin{equation}
  \label{eq:WL_ac2}
  \begin{aligned}
 &   \hat{A}_{E\mu}(X_{E}) dX_{E\mu}
=\hat{A}_{\mu}(X)  dX^\mu|_{r_\para\to -i r_\para}\,.
  \end{aligned}
\end{equation}
It then follows that
\begin{equation}
  \label{eq:WL_ac}
\hat{\W}_E^{\,(T)}(r_\para,\vec r_\perp)   =
\hat{\W}_M^{\,(T)}(-ir_\para,\vec r_\perp)\,, 
\end{equation}
again to be understood in the weak sense.

At a certain stage of the calculation we will need Euclidean
Wilson-loop operators running along the same paths $\C_E(0,R_{E},T)$
appearing in Eq.~\eqref{eq:WL_notation}, but corresponding to a
different choice of the Euclidean ``time'' direction, i.e., obeying a
different time-ordering. These operators will be denoted by
\begin{equation}
  \label{eq:WL_othert}
\hat{\W}_{E*}^{\,(T)}(r_\para,\vec r_\perp)   = 
\f{1}{N_c} \tr {\rm T_1P} \exp\left\{
-ig\oint_{\C_E(0,R_{E},T)} \hat A_{E\mu}(X_E) d X_{E\mu} \right\}\,,
\end{equation}
where
${\rm T_1}$ denotes time-ordering along direction 1, i.e. (for $T\to
\infty$), along the ``long'' side of the loops.

Let us finally notice that time- and path-ordering can be disentangled
using the continuous-product representation for the path-ordered
exponential. In the Minkowskian case, Eq.~\eqref{eq:wlE_op_M},
parameterising the path as $X(\lambda)$, with $\lambda\in[0,1]$ and
$X(0)=X(1)$, one has 
\begin{equation}
  \label{eq:cont_prod}
\hat{\W}_{M}[\C]
= \lim_{N\to\infty} \sum_{j_1,\ldots, j_N=1}^{N_c}\T\left\{\prod_{k=0}^{N-1}
  \hat{w}^{(N)}_{j_k j_{k-1}}(k)\right\}\,, 
\end{equation}
where $j_k$ are group indices, 
\begin{equation}
  \label{eq:cont_prod2}
\hat{w}^{(N)}(k) =  \mathbf{1}
-\f{ig}{N}\hat A_{\mu}\left(X_k\right)
\dot{X}_k^{\mu} \,,
\qquad
X_k=X\left(\tf{1}{N}\left(k+\tf{1}{2}\right)\right)\,,\quad
\dot{X}_k=\f{dX}{d\lambda}\left(\tf{1}{N}\left(k+\tf{1}{2}\right)\right)\,,
\end{equation}
with $\mathbf{1}$ the group identity, and $\hat{w}^{(N)}_{j_k
  j_{k-1}}(k)$ in Eq.~\eqref{eq:cont_prod} are ordered according to
$X_k^0$. Similar representations hold for the Euclidean Wilson loop
operators defined in Eqs.~\eqref{eq:wlE_op} and \eqref{eq:WL_othert}.

\subsection{Complete set of states}
\label{subsec:not}

The approach followed in this paper to determine the large-distance
behaviour of the static dipole-dipole potential is based on the
insertion of a complete set of states in a certain Wilson-loop
correlation function. We use the complete set of asymptotic ``in''
states, characterised by their particle content,
and by the momenta and third component of the spins of the
particles. We define here the setup in full generality, so that the
results obtained in this paper can be applied to a wide class of gauge
theories. 

Let the spectrum of asymptotic states contain $n_{\rm sp}$ different
species of stable particles, characterised by their mass $m_{(s)}$ and
spin $s_{(s)}$, with $s\in
\{1,\ldots,n_{\rm  sp}\}$. The particle content of a state is
specified by the string $\alpha=\{ N_1,N_2,\ldots,N_{n_{\rm
    sp}}\}$ of the occupation numbers $N_s=N_s(\alpha)$. 
For the vacuum $N_s=0~\forall s$ we use the notation 
$\alpha=\emptyset$. 
Particles are labelled by a double index $\is$, taking values in the
index space 
$S=\{\is~|~ \is\in \mathbb{N},\,s=1,\ldots,n_{\rm sp}\}$. 
For a given particle content $\alpha$, indices run over
the set
\begin{equation}
  \label{eq:Salpha}
   S_\alpha =  \{ \is\in S ~|~ 1\le \is \le
N_s(\alpha)\,,~N_s(\alpha)\ne 0  \}\,;
\end{equation}
the total number of particles is ${\cal N}_\alpha=\sum_s
N_s(\alpha)$. 
The momenta, $\vec p_{\is}$, and the third component of the spins,
$s_{3\is}$, of all the particles
in a state are denoted collectively as $\Omega_{S_\alpha}$,
where for a general $A\subseteq S$
\begin{equation}
  \label{eq:OmegaA}
  \Omega_{A}  =
\{(\vec p_{\is},s_{3\is}) ~|~ \is\in A\}\,. 
\end{equation}
A state is completely specified by $\alpha$ and
$\Omega_{S_\alpha}$, and will be denoted as follows,
\begin{equation}
  \label{eq:state}
   | \Omega_{S_\alpha}  \ra \equiv 
| \cup_{\is\in S_\alpha}\{\vec p_{\is},s_{3\is}\}
 ~;~{\rm in}\ra\,,
\end{equation}
where the right-hand side stands for the ``in'' state with the
appropriate particle content. Such a state transforms under translations
and Lorentz transformations as the properly (anti)sym\-me\-trised
tensor product of the corresponding one-particle states, and obeys the  
usual relativistic normalisation. 
For off-shell momenta, we denote by 
$ \tilde\Omega_{A}  = \{(p_{\is},s_{3\is}) ~|~ \is\in A\}
$ the collection of four-momenta and spins.
The total energy of a state is denoted as $E(\Omega_{S_\alpha})$,
where for any $A\subseteq S$
\begin{equation}
  \label{eq:energy}
E({\Omega_A}) = \sum_{\is\in A} \varepsilon_{\is}\,, \qquad
\varepsilon_{\is} = \sqrt{\vec p_{\is}^{\,2} +   m_{(s)}^2}\,.  
\end{equation}
Finally, completeness is expressed as
\begin{equation}
  \label{eq:completeness}
1= \sum_\alpha \f{1}{\prod_s N_s(\alpha)!}\int d\Omega_{S_\alpha}\,
 | \Omega_{S_\alpha} 
\ra \la \Omega_{S_\alpha}  |  \,,
\end{equation}
where for any $A\subseteq S$
\begin{equation}
  \label{eq:phasespace}
\int d\Omega_{A} = \int \prod_{\is\in A} \left[
\f{d^3p_{\is}}{(2\pi)^3
  2\varepsilon_{\is}}\sum_{s_{3\is}=-s_{(s)}}^{s_{(s)}}\right]\,.
\end{equation}
In the following we will also use the notation
\begin{equation}
  \label{eq:doubleangbrack}
\lla f(\Omega_{A})\rra_{\Omega_A;\,b} = \int 
d\Omega_A\,e^{-bE(\Omega_A) } f(\Omega_A)\,.  
\end{equation}

\begin{figure}[t]
  \centering
  \includegraphics[width=0.57\textwidth]{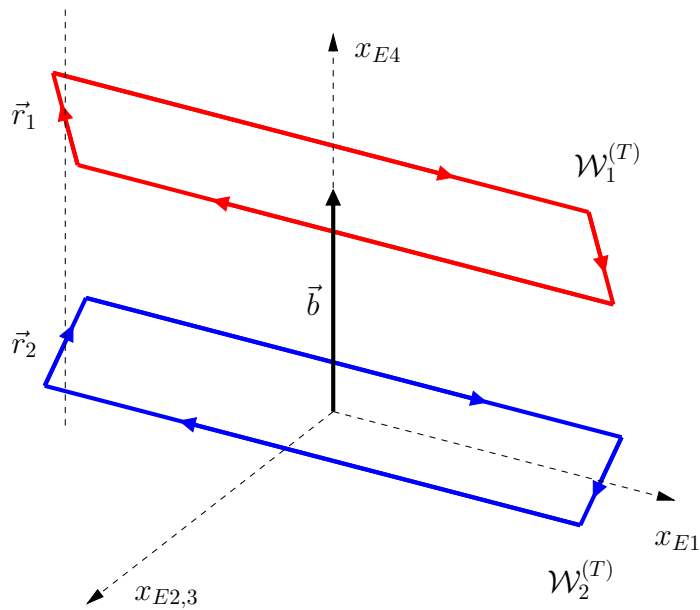}
  \caption{The relevant Euclidean Wilson loops.}
  \label{fig:1}
\end{figure}

\section{Dipole-dipole potential from a sum over states}
\label{sec:wlc}

The potential ${V}_{dd}$ between two static dipoles of size
$\vec{r}_{1,2}$, with centers separated by $\vec{b}$, is obtained from
the correlation function of two rectangular $T\times |\vec r_{1,2}|$
Euclidean Wilson loops, in the limit of large $T$: 
\begin{equation}
  \label{eq:ddpot1}
  e^{-T{V}_{dd}} \mathop =_{T\to\infty}
\f{\la \W_1^{(T)} \W_2^{(T)} \ra_E}{\la \W_1^{(T)}
  \ra_E \la \W_2^{(T)} \ra_E}\,,
\end{equation}
where $\W_{1,2}^{(T)}=\W_E[\C_{1,2}]$ for properly chosen paths
$\C_{1,2}$ [see Eq.~\eqref{eq:wlE_FI}], and $\la\ldots\ra_E$ denotes
the expectation value in the Euclidean functional-integral sense. 
Without referring to a specific Euclidean coordinate system (for
reasons that will be apparent shortly), the Wilson-loop configuration
can be described as follows. The size and the relative orientation of
the ``short'' sides of the loops and of the separation between their
centers coincide with the size and the relative orientation of
$\vec{r}_{1}$, $\vec{r}_{2}$, and $\vec{b}$. The ``long'' sides of the
two loops have length $T$, lie in the orthogonal complement of the
vector subspace determined by $\vec{r}_{1}$, $\vec{r}_{2}$, and
$\vec{b}$, and are parallel. 
In a nutshell, our approach to the determination of the large-distance
behaviour of the dipole-dipole potential consists of going over to the
operator formalism and inserting a complete set of states between the
loops. Before setting up the calculation in full detail, let us
briefly discuss the potentially confusing issue of Euclidean ``time''.

Usually, the long sides of the loops are taken to be parallel to the
Euclidean ``time'' direction, so that the loops describe the evolution
of the dipoles over an amount $T$ of Euclidean ``time'', which is
eventually taken to infinity. However, the notion of Euclidean
``time'' is well defined only after setting up the Hamiltonian
formulation of the theory, while in the Lagrangian formulation
employed in the functional-integral formalism the direction of
``time'' can be chosen arbitrarily, thanks to the $O(4)$ 
invariance of the Euclidean theory. In our approach we exploit this
arbitrariness, and we use two different choices of ``time'' at
different stages of the 
calculation. For our purposes, it is convenient at first to take
``time'' along the separation between the centers of the loops. In
this way, after going over to the operator formalism, one can extract
the large-distance behaviour of the loop-loop correlator in the usual
way, by inserting a complete set of states between the Wilson-loop
operators. Schematically,
\begin{equation}
  \label{eq:sketch1}
\la \W_1^{(T)} \W_2^{(T)} \ra_E  
= \sum_n e^{-E_n |\vec b|}\la 0 | \hat\W_E^{(T)}(r_{1\parallel}, \vec
r_{1\perp})|n\ra\la n|   \hat\W_E^{(T)}(r_{2\parallel}, \vec r_{2\perp})
| 0\ra\,,
\end{equation}
where $\W_E^{(T)}(r_{\parallel}, \vec r_{\perp})$ denotes a
Wilson loop centered at the origin, spanning a temporal
interval of size $|r_{\parallel}|$, and with a long spatial side of size
$T$ [for more details see
Eqs.~\eqref{eq:WL_path_b}--\eqref{eq:WL_notation}, 
and below in this Section], and a caret denotes the
corresponding operator. Here $|n\ra$ denotes an energy eigenstate.
The large-distance behaviour of the correlator is then obtained from
the contributions of the lightest states to Eq.~\eqref{eq:sketch1}. 
To determine the dipole-dipole potential one has to further take
the limit $T\to\infty$, which affects the Wilson-loop matrix elements
appearing in Eq.~\eqref{eq:sketch1}. As we show below in Section
\ref{sec:me}, these matrix elements can be related to the correlation
function (again in the sense of the Euclidean functional integral) of a
Wilson loop with appropriate Euclidean interpolating fields,
 corresponding to the particles appearing in the state
$|n\ra$. To study these correlation functions, it is useful to go over
again to the operator formalism, but with choosing now the ``time''
direction along the long side of the loop:
\begin{equation}
  \label{eq:sketch2}
  \la 0 | \hat\W_E^{(T)}(r_\parallel, \vec r_\perp)|n\ra \longrightarrow
\la \W_E^{(T)}(r_\parallel, \vec r_\perp) \prod_i\phi_{Ei}(x_i)\ra_E
= \la 0| \T_1\left\{\hat\W_{E*}^{(T)}(r_\parallel, \vec r_\perp)
  \prod_i\hat\phi_{Ei}(x_i)\right\}| 0\ra\,, 
\end{equation}
where $\phi_{Ei}(x_i)$ are the Euclidean interpolating fields, and $\T_1$
denotes time-ordering of the operators along the direction of the long 
side of the loop. Here the Wilson-loop operator is denoted by
$\hat\W_{E*}^{(T)}$, to make it explicit that a different
time-ordering is used [see Eq.~\eqref{eq:WL_othert}]: although
$\hat\W_{E}^{(T)}$ and $\hat\W_{E*}^{(T)}$ correspond to the same
Euclidean path, they are in effect different operators.
From the representation Eq.~\eqref{eq:sketch2} we can then
establish the relevant properties of the matrix element in the
large-$T$ limit, and by taking $T\to\infty$ we can finally derive the
dipole-dipole potential at large distances.  
We want to stress the fact that it is perfectly legitimate to
use different choices for the ``time'' direction to recast the
same (functional-integral) correlation function in the operator
formalism in different ways, in order to study different aspects of
said correlation function, as long as these choices are used
consistently. In our case, different choices for the ``time'' direction
are made in the study of different correlation functions, namely the
loop-loop [Eq.~\eqref{eq:sketch1}] and the loop-interpolating fields
[Eq.~\eqref{eq:sketch2}] correlation functions, so that no 
inconsistency can arise. 
We also want to remark that the physical, Minkowskian time plays no
role in our calculation, which, starting from Eq.~\eqref{eq:ddpot1},
can in principle be performed entirely in Euclidean
space.\footnote{The Minkowskian Wilson loops used in 
  Section \ref{sec:me} have to be regarded simply as a useful
  mathematical device: those loops have in fact no relation with the
  physical process of static dipoles evolving over a large physical
  time.}  

Let us now return to Eq.~\eqref{eq:ddpot1} and proceed in a more
detailed fashion. As we have said above, the $O(4)$ invariance of the
Euclidean theory allows us to choose freely the global orientation of
the Wilson-loop configuration. For our purposes, it is convenient to
choose $\C_{1,2}$ as follows (see Fig.~\ref{fig:1}), 
\begin{equation}
  \label{eq:WL_path_b_0}
  \begin{aligned}
  \C_1 &=\C_E(z_E,R_{E1},T)\,, &&&
  \C_2&=\C_E(0,R_{E2},T)\,,
  \end{aligned}
\end{equation}
where the paths $\C_E(z_E,R_{E},T)$ have been defined in
Eq.~\eqref{eq:WL_path_b}, and
\begin{equation}
  \label{eq:WL_path3_bis}
  R_{E1,2} = (0,\vec r_{1,2\perp},r_{1,2\para})=(0,\vec r_{1,2})\,, 
  \qquad z_E=(0,\vec b_\perp,b_\para)=(0,\vec{b})\,.
\end{equation}
The Euclidean $O(4)$ invariance further allows us to set $\vec
b_\perp=0$ and $b_\para = |\vec b\,| \ge 0$ with no loss of generality. 
We can thus work in this coordinate frame, and write 
$V_{dd}=V_{dd}(b;r_{1\para},\vec{r}_{1\perp},r_{2\para},\vec{r}_{2\perp})$, with
$b\equiv |\vec b\,|$, so that
\begin{equation}
  \label{eq:ddpot2_0}
{\cal G}^{(T)}(b;r_{1\para},\vec{r}_{1\perp},r_{2\para},\vec{r}_{2\perp})\equiv
\f{\la \W_1^{(T)} \W_2^{(T)} \ra_E}{\la \W_1^{(T)} \ra_E \la
  \W_2^{(T)} \ra_E} =
e^{-TV_{dd}(b;r_{1\para},\vec{r}_{1\perp},r_{2\para},\vec{r}_{2\perp})
+ o(T)} \,.
\end{equation}
In the operator formalism, this correlation function reads
\begin{equation}
  \label{eq:ddpot2}
{\cal G}^{(T)}(b;r_{1\para},\vec{r}_{1\perp},r_{2\para},\vec{r}_{2\perp})=
\f{\la 0 | \T\left\{\hat\W_1^{(T)} \hat\W_2^{(T)}\right\}|0 \ra}{\la 0|
  \hat\W_1^{(T)} |0\ra \la 0| \hat\W_2^{(T)} |0 \ra}\,,
\end{equation}
where $\hat{\W}_{1,2}^{\,(T)}=\hat{\W}_E[\C_{1,2}]$ [see
Eq.~\eqref{eq:wlE_op}]. For loops that do not overlap in the
``temporal'' direction, i.e., for $b> |r_{1\para}| + |r_{2\para}|$, the
$\T$-ordering sign can be omitted, and so one can insert a complete
set of states between the loops. Since in this paper we are interested
in the asymptotic large-distance behaviour of the potential, we will
restrict to this case, 
without loss of generality. Exploiting ``time''-translation
invariance, we can write 
\begin{equation}
  \label{eq:ddpot3}
  \begin{aligned}
&{\cal G}^{(T)}(b;r_{1\para},\vec{r}_{1\perp},r_{2\para},\vec{r}_{2\perp})=
\sum_\alpha \f{1}{\prod_s N_s(\alpha)!}\,
G^{(T)}_{S_\alpha}(b;r_{1\para},\vec{r}_{1\perp},r_{2\para},\vec{r}_{2\perp}) 
\,,
  \end{aligned}
\end{equation}
where 
\begin{equation}
  \label{eq:ddpot_MM2}
  \begin{aligned}
    G^{(T)}_{S_\alpha}(b;r_{1\para},\vec{r}_{1\perp},r_{2\para},\vec{r}_{2\perp}) 
    &=\lella
    M^{\,(T)}(\Omega_{S_\alpha};r_{1\para},\vec r_{1\perp})
   \bar{M}^{\,(T)}(\Omega_{S_\alpha};r_{2\para},\vec r_{2\perp})
   \rirra_{\Omega_{S_\alpha};\,b}
   \\ 
   &= 
\int d\Omega_{S_\alpha}\,
e^{-bE(\Omega_{S_\alpha})} 
    M^{\,(T)}(\Omega_{S_\alpha};r_{1\para},\vec r_{1\perp})
   \bar{M}^{\,(T)}(\Omega_{S_\alpha};r_{2\para},\vec r_{2\perp})
\,,
  \end{aligned}
\end{equation}
and we have denoted as follows the relevant Wilson-loop matrix
elements, 
\begin{equation}
  \label{eq:ddpot_MM}
  \begin{aligned}
   M^{\,(T)}(\Omega_{S_\alpha};r_\para,\vec r_\perp) & \equiv
\f{\la 0 |\hat{\W}_E^{\,(T)}(r_\para,\vec r_\perp)
    | \Omega_{S_\alpha} \ra }
{\la 0 |\hat{\W}_E^{\,(T)}(r_\para,\vec r_\perp)
    | 0 \ra}\,, &&& 
       \bar M^{\,(T)}(\Omega_{S_\alpha};r_\para,\vec r_\perp) &\equiv 
\f{\la \Omega_{S_\alpha} |\hat{\W}_E^{\,(T)}(r_\para,\vec r_\perp)
    | 0 \ra }
{\la 0 |\hat{\W}_E^{\,(T)}(r_\para,\vec r_\perp)
    | 0 \ra}\,,
  \end{aligned}
\end{equation}
where $\hat{\W}_E^{\,(T)}(r_\para,\vec r_\perp)$ is computed on the path
$\C_E(0,R_{E},T)$ and has been defined in
Eq.~\eqref{eq:WL_notation}. Notice that for the vacuum state $
G^{(T)}_{S_\emptyset}=1$.  
The two quantities $M^{\,(T)}$ and $\bar M^{\,(T)}$ can be treated at
once by noticing that under Hermitian conjugation
\begin{equation}
  \label{eq:herm_conj_wl}
  \left[\hat{\W}_E^{\,(T)}(r_\para,\vec r_\perp)\right]^\dag 
= \hat{\W}_E^{\,(T)}(r_\para,-\vec r_\perp)
\,,
\end{equation}
and so it is straightforward to show that  
\begin{equation}
  \label{eq:wilson_bar_lsz3}
  \begin{aligned}
   \bar M^{\,(T)}(\Omega_{S_\alpha};r_\para,\vec r_\perp) &= 
   \left[ M^{\,(T)}(\Omega_{S_\alpha};r_\para,-\vec r_\perp)\right]^*\,.
  \end{aligned}
\end{equation}
In the remainder of this Section we show how the expression
Eq.~\eqref{eq:ddpot3} for the Wilson-loop correlator exponentiates to the
form given in Eq.~\eqref{eq:ddpot2_0}, with the right $T$-dependence
in the large-$T$ limit. The strategy we pursue is the following. We
first derive, in Subsection \ref{sec:me}, a Euclidean
Lehmann--Symanzik--Zimmermann (LSZ)~\cite{LSZ1,LSZ2} 
representation for the matrix elements, Eq.~\eqref{eq:ddpot_MM}, and
from this we obtain, in Subsection \ref{sec:cd}, a decomposition of
the matrix elements in connected components, with each component
describing, loosely speaking, the interaction of an isolated subset of
particles with the loop. This decomposition allows us to prove the
exponentiation of Eq.~\eqref{eq:ddpot3}, and finally to establish that
the correlator exhibits the correct dependence on $T$, in Subsection
\ref{sec:Tdep}, where the final expression for the dipole-dipole
potential is also reported.\footnote{We notice, incidentally, that the
  exponentiation of Eq.~\eqref{eq:ddpot3} could be formally obtained in a
  straightforward way by means of the moments-cumulant
  theorem. However, this would tell us nothing about the properties of
  the exponent, so that we could not prove that the correlator has the 
  right $T$-dependence.}

\subsection{Euclidean LSZ representation for the matrix elements}
\label{sec:me}

The relevant Euclidean matrix elements
$M^{\,(T)}(\Omega_{S_\alpha};r_\para,\vec r_\perp)$ 
are related to the analogous matrix elements for the
Minkowskian Wilson loop,
\begin{equation}
  \label{eq:wilson_lsz2}
   M_M^{\,(T)}(\Omega_{S_\alpha};r_\para,\vec r_\perp) \equiv 
\f{\la 0 |\hat{\W}_M^{\,(T)}(r_\para,\vec r_\perp)
    |  \Omega_{S_\alpha} \ra }
{\la 0 |\hat{\W}_M^{\,(T)}(r_\para,\vec r_\perp)
    | 0 \ra }\,,
\end{equation}
by means of analytic continuation [see Eq.~\eqref{eq:WL_ac}],
\begin{equation}
\label{eq:wilson_lsz3}
   M^{\,(T)}(\Omega_{S_\alpha};r_\para,\vec r_\perp) = 
   M_M^{\,(T)}(\Omega_{S_\alpha};-ir_\para,\vec r_\perp)\,.
\end{equation}
Although the physical quantities entering the dipole-dipole potential
are the Euclidean matrix elements $M^{\,(T)}$, in order to recast them
into a LSZ-like expression it is convenient to work initially with
$M_M^{\,(T)}$. The quantity $M_M^{\,(T)}$ admits in fact a
straightforward LSZ reduction~\cite{LSZ1,LSZ2}, which can be written
in the following compact form,\footnote{The derivation of
  Eqs.~\eqref{eq:wilson_lsz4} and \eqref{eq:wilson_lsz4_bis} follows
  the usual LSZ procedure, the only nontrivial point being the
  definition of a time-ordered product involving local fields and the
  nonlocal Wilson-loop operator. This can however be easily obtained
  by using the continuous-product representation of the Wilson loop,
  Eq.~\eqref{eq:cont_prod}, which allows us to write
  \begin{equation*}
    \T\left\{\hat{\W}_M[\C] \prod_{i} \hat\phi_i(x_i)  \right\}
    = \lim_{N\to\infty} \sum_{j_1,\ldots,
      j_N=1}^{N_c}\T\left\{\prod_{k=0}^{N-1} 
      \hat{w}^{(N)}_{j_k j_{k-1}}(k)\prod_i \hat\phi_i(x_i)\right\}\,, 
  \end{equation*}
  for a general path $\C$, and for a general set of local fields
  $\hat\phi_i(x_i)$.}
\begin{equation}
\label{eq:wilson_lsz4}
\begin{aligned}
  M_M^{\,(T)}(\Omega_{S_\alpha};r_\para,\vec r_\perp) &= 
{\rm Lim}_{S_\alpha}
\Pi(\tilde\Omega_{S_\alpha})\, L_{M}^{\,(T)}(P^0_{S_\alpha},\vec P_{S_\alpha}
  ;r_\para,\vec r_\perp) \,,\\
L_{M}^{\,(T)}(P^0_{S_\alpha},\vec P_{S_\alpha};r_\para,\vec r_\perp) &\equiv
    \int dX_{{S}_\alpha} \,
 e^{-iP_{S_\alpha}\cdot X_{S_\alpha}} \C^{(T)}_M(X^0_{S_\alpha},\vec
 X_{S_\alpha};r_\para,\vec r_\perp)\,, 
\end{aligned}
\end{equation}
where $\C^{(T)}_M$ and $\Pi$ are defined as follows
\begin{equation}
  \label{eq:wilson_lsz4_bis}
  \begin{aligned}
    \C^{(T)}_M(X^0_{S_\alpha},\vec
 X_{S_\alpha};r_\para,\vec r_\perp) &\equiv
  \f{\la 0 |\T\left\{
    \hat{\W}_M^{\,(T)}(r_\para,\vec r_\perp) 
\prod_{\is \in {S}_\alpha} \hat\Phi^{(s)}(x^0_{\is},\vec x_{\is}) 
  \right\}| 0 \ra }{\la 0 |\hat{\W}_M^{\,(T)}(r_\para,\vec r_\perp)
    | 0 \ra } \\
  &=  
  \f{\left\la {\W}_M^{\,(T)}(r_\para,\vec r_\perp) 
\prod_{\is \in {S}_\alpha} \Phi^{(s)}(x^0_{\is},\vec x_{\is}) 
 \right\ra_M}{\left\la {\W}_M^{\,(T)}(r_\para,\vec r_\perp)
    \right\ra_M } \,,\\
  \Pi(\tilde\Omega_{S_\alpha})
&\equiv \prod_{\is \in {S}_\alpha} \pi^{(s)}( p_{\is},s_{3\is})
\,.
  \end{aligned}
\end{equation}
Here we have introduced some notation that we now explain. $\Pi$
denotes the product of the ``projectors'' on the appropriate 
particle poles and spin components: for example, for a scalar particle
of mass $m$, $\pi^{(0)}(p)=p^2-m^2$;  
for a spin-$\f{1}{2}$ fermion,  
$\pi^{(\f{1}{2})}(p,s_3)=(\slap - m)u_{s_3}(\vec p\,)=(p^2-m^2)(\slap +
m)^{-1}u_{s_3}(\vec p\,)$, and so on. 
Both $L_{M}^{\,(T)}$ and $\Pi$
(may) carry Lorentz indices, appropriately contracted in the product,
and are first evaluated off-shell; the on-shell limit, denoted with
\begin{equation}
  \label{eq:onshelllim}
{\rm Lim}_{S_\alpha}  
= \prod_{\is\in {S_\alpha}} \lim_{p_{\is}^2 \to m_{(s)}^2}\,,  
\end{equation}
is taken after computing the product. 
The operators $\hat\Phi^{(s)}(x^0_{\is},\vec x_{\is})$ are the
appropriate local interpolating fields for particles of type $s$,
normalised to have free-field one-particle matrix elements, i.e., 
the renormalisation constants required in the LSZ formulae have been
absorbed in their definition, so that $\hat\Phi^{(s)}$ are
renormalised fields.   
We have denoted collectively with $P_{S_\alpha}=(P^0_{S_\alpha},\vec
P_{S_\alpha})$ the four-momenta of the particles, and the temporal and
spatial components thereof. A similar collective notation, 
$X_{S_\alpha}=(X^0_{S_\alpha},\vec X_{S_\alpha})$, has been used for
the coordinates of the local operators, 
and for the
corresponding integration measure, $dX_{{S}_\alpha}$. 
In the following, when
there is no need to distinguish between temporal and spatial
components, we do not write them as separate arguments.
Moreover,
 $ P_{S_\alpha}\cdot  X_{S_\alpha} =
 \sum_{\is\in S_\alpha} p_{\is\mu} x^\mu_{\is} 
 $. 
Finally, in the second line of Eq.~\eqref{eq:wilson_lsz4_bis} we have
used the functional-integral representation for time-ordered vacuum expectation
values, denoting with $\la\ldots\ra_M$ the expectation value
in the sense of the Minkowskian functional integral. 

The next step is to Wick-rotate $L_{M}$ to Euclidean
space. By means of a simple change of variables, one shows that
\begin{equation}
  \label{eq:wick_rot}
  \begin{aligned}
  L_{M}^{\,(T)}(\xi^{-1} P^0_{S_\alpha},\vec P_{S_\alpha};\xi r_\para,\vec r_\perp)
 &=
\xi^{{\cal N}_\alpha} \int
dX_{S_\alpha} e^{-i
P_{{S}_\alpha}\cdot X_{{S}_\alpha}
 } \C^{(T)}_M(\xi X^0_{S_\alpha},\vec
 X_{S_\alpha};\xi r_\para,\vec r_\perp)
\,.
\end{aligned}
\end{equation}
By sending $\xi \to e^{-i\f{\pi}{2}}$ we then obtain
\begin{equation}
  \label{eq:wick_rot2}
  \begin{aligned}
    L_{M}^{\,(T)}(e^{i\f{\pi}{2}} P^0_{S_\alpha},\vec
P_{S_\alpha};e^{-i\f{\pi}{2}} r_\para,\vec r_\perp) 
=
(-i)^{{\cal N}_\alpha} 
L_{E}^{\,(T)}(-P^0_{S_\alpha},\vec
P_{S_\alpha}; r_\para,\vec r_\perp)  \,, 
  \end{aligned}
\end{equation}
where we have introduced the Euclidean quantity
\begin{equation}
  \label{eq:wick_rot2_bis}
  \begin{aligned}
L_{E}^{\,(T)}(P_{E S_\alpha 4},\vec
P_{E S_\alpha}; r_\para,\vec r_\perp)  &\equiv 
    \int dX_{E\,S_\alpha} e^{i P_{E{S}_\alpha } \cdot  X_{E\,{S}_\alpha }}
\C^{(T)}_E(\vec X_{E\, S_\alpha}, X_{E\,  S_\alpha \,4},; r_\para,\vec r_\perp)\,,\\
 \C^{(T)}_E(\vec X_{E\, S_\alpha}, X_{E\,  S_\alpha \,4},; r_\para,\vec
r_\perp) & \equiv
\f{  \la 0 |\T\left\{
    \hat{\W}_E^{\,(T)}(r_\para,\vec r_\perp) 
\prod_{\is \in {S}_\alpha} \hat\Phi^{(s)}_E(\vec
x_{E\is},x_{E\is 4}) 
  \right\}| 0 \ra }
{\la 0 |
    \hat{\W}_E^{\,(T)}(r_\para,\vec r_\perp) | 0 \ra
}  \\ &=
  \f{\left\la {\W}_E^{\,(T)}(r_\para,\vec r_\perp) 
\prod_{\is \in {S}_\alpha} \Phi^{(s)}_E(\vec x_{E\is},x_{E\is 4}) 
 \right\ra_E}{\left\la {\W}_E^{\,(T)}(r_\para,\vec r_\perp)
    \right\ra_E } \,,
  \end{aligned}
\end{equation}
where $P_{E \,S_\alpha}=(\vec P_{E \,S_\alpha},P_{E \,S_\alpha\,4})$
denotes collectively the Euclidean four-momenta $p_{E\,\is}$,
$X_{E\,S_\alpha}\!=\!(\vec 
X_{E\,S_\alpha}, X_{E\,S_\alpha\,4})$ the coordinates of the local
operators, $dX_{E\,S_\alpha}$ the
corresponding integration measure, and 
 $ P_{E\,S_\alpha}\cdot  X_{E\,S_\alpha} =
 \sum_{\is\in S_\alpha} p_{E\,\is\mu} x_{E\,\is\mu} $. 
The Euclidean Wilson loop $\hat{\W}_E^{\,(T)}$ has been
defined in Eq.~\eqref{eq:wlE_op}, and 
$\hat\Phi_{E}^{(s)}$ are now the appropriate local functionals of the
Euclidean fields. In the third line we have made contact with the
Euclidean functional-integral formalism. 

Inverting the analytic-continuation relation we find\footnote{The
  notation $e^{-i\f{\pi}{2}}(-P^0_{S_\alpha})$ indicates that to
  obtain the correlator at positive (off-shell) energies $p^0_{\is}$
  one starts from negative $p_{E\,\is\,4}=-p^0_{\is}$, and then
  rotates clockwise in the complex $p_{E\,\is\,4}$ plane.}
\begin{equation}
  \label{eq:wick_rot3}
  \begin{aligned}
&    L_{M}^{\,(T)}(P^0_{S_\alpha},\vec P_{S_\alpha}
; r_\para,\vec r_\perp) =
(-i)^{{\cal N}_\alpha} 
L_{E}^{\,(T)}(e^{-i\f{\pi}{2}}(-P^0_{S_\alpha}),\vec P_{S_\alpha}
;e^{i\f{\pi}{2}} r_\para,\vec r_\perp) \,,
 \\
&    L_{M}^{\,(T)}(P^0_{S_\alpha},\vec P_{S_\alpha}
;-i r_\para,\vec r_\perp) =
(-i)^{{\cal N}_\alpha}  
L_{E}^{\,(T)}(e^{-i\f{\pi}{2}}(-P^0_{S_\alpha}),\vec P_{S_\alpha}
; r_\para,\vec r_\perp)   \,.
  \end{aligned}
\end{equation}
Summarising, $M^{\,(T)}$ is obtained by first computing
$L_{E}^{\,(T)}$ for real arguments, then performing the Wick rotation to  
obtain $L_{M}^{\,(T)}$ (with real arguments), taking the momenta
on-shell and finally analytically continuing $r_\para\to -i r_\para$. 
However, the on-shell projection and the last analytic continuation
$r_\para\to -i r_\para$ should not interfere. If it is so, then
\begin{equation}
  \label{eq:lsz_proj_2}
     M^{\,(T)}(\Omega_{S_\alpha}
;r_\para,\vec r_\perp) = 
{\rm Lim}_{S_\alpha}
    \Pi(\tilde\Omega_{S_\alpha})
 (-i)^{{\cal N}_\alpha} 
    L_{E}^{\,(T)}(e^{-i\f{\pi}{2}}(-P^0_{S_\alpha}),\vec P_{S_\alpha}
;r_\para,\vec r_\perp)\,, 
\end{equation}
and we can follow a simpler route: after computing
$L_{E\alpha}^{\,(T)}$ for real arguments, we
perform the partial Wick rotation $L_{E\alpha}^{\,(T)}(
e^{-i\f{\pi}{2}}(-P^0_{S_\alpha}),\vec P_{S_\alpha}
;  r_\para,\vec r_\perp)$, and finally take the momenta on-shell.

Let us finally notice that the correlator $\C^{(T)}_E$ in
Eq.~\eqref{eq:wick_rot2_bis} is a renormalised quantity. Indeed, we
are working with renormalised interpolating fields [see discussion
after Eq.~\eqref{eq:onshelllim}], and moreover the Euclidean Wilson
loop enters $\C^{(T)}_E$ through the combination
$\W^{(T)}_E/\la\W^{(T)}_E\ra_E$, which is a renormalisation-invariant
quantity since $\W^{(T)}_E$ renormalises
multiplicatively~\cite{DV,BNS}. As a consequence, the matrix elements
$M^{\,(T)}$, Eq.~\eqref{eq:ddpot_MM} [as well as the Minkowskian  
matrix elements $M_M^{\,(T)}$, Eq.~\eqref{eq:wilson_lsz2}], are 
renormalised (and renormalisation-invariant) quantities.

\subsection{Cluster decomposition of the matrix elements}
\label{sec:cd}

The point in relating $M^{\,(T)}$ with the purely
Euclidean quantity $L_{E}^{\,(T)}$ is that the latter admits a neat
cluster decomposition. Furthermore, as the Euclidean functional
integral admits a nonperturbative definition through the lattice
discretisation, we can perform the formal manipulations rather safely.  
To compute the correlation function $\C^{(T)}_E$,
Eq.~\eqref{eq:wick_rot2_bis}, 
we can exploit once again the $O(4)$ invariance of the Euclidean
theory, and choose the ``time'' direction as we please. For our
purposes, it is convenient to now take ``time'' along direction 1,
i.e., the direction of the ``long'' side of the loop. Reverting to the
operator formalism with this choice for the ``time'' direction, we
write 
\begin{equation}
  \label{eq:clust1bis2}
  \begin{aligned}
    \C_E^{(T)}(X_{E\,S_\alpha};r_\para,\vec r_\perp)  &= 
\f{  \la 0 |\T_1 \left\{
    \hat{\W}_{E*}^{\,(T)}(r_\para,\vec r_\perp) 
\hat\Op_{S_\alpha}(X_{E\,{S}_\alpha})  
  \right\}| 0 \ra }
{\la 0 |
    \hat{\W}_{E*}^{\,(T)}(r_\para,\vec r_\perp) | 0 \ra
} \,,\\
\hat\Op_{S_\alpha}(X_{E\,{S}_\alpha})    &\equiv 
\prod_{\is \in S_\alpha} \hat\Phi^{(s)}_E(\vec
x_{E\,\is},x_{E\,\is\, 4}) \,, 
  \end{aligned}
\end{equation}
where $\hat{\W}_{E*}^{\,(T)}$ has been defined in
Eq.~\eqref{eq:WL_othert}, and $\T_1$ denotes time-ordering along
direction 1. As $\C_E^{(T)}$ is gauge invariant, we can work in 
the temporal gauge where the long sides of the loop are trivial.
With this choice of time-ordering and in this gauge, the
Wilson-loop operator can be expressed in terms of the following
Wilson-line operator, 
\begin{equation}
  \label{eq:wlE_bis}
  \hat{W}_E(R_E) = {\rm P} \exp\left\{
-ig\int_{-\f{1}{2}}^{+\f{1}{2}}d\lambda\, \hat A_{E\mu}(\lambda R_E)
R_{E\mu} \right\} \,, \qquad   \hat{W}_E(-R_E)=  \hat{W}_E(R_E)^\dag\,,
\end{equation}
where the time-ordering symbol has been dropped, since only gauge
fields at $x_{E1}=0$ appear.
In terms of $\hat{W}_E$, the Wilson-loop operator reads 
\begin{equation}
  \label{eq:clust2}
  \begin{aligned}
    \hat{\W}_{E*}^{\,(T)}(r_\para,\vec r_\perp) 
&=    {\f{1}{N_c}}\tr \left\{
e^{ \hat H\f{T}{2}}\hat W_E(R_E)^\dag e^{- \hat   H \f{T}{2}} e^{- \hat
  H \f{T}{2}}\hat W_E(R_E) e^{ \hat H\f{T}{2}} 
\right\}\,,
  \end{aligned}
\end{equation}
with $\hat{H}$ the Hamiltonian operator. Since we are ultimately
interested in the limit $T\to\infty$, we consider only the case when 
$T/2>|x_{E\,\is\, 1}|$ $\forall \is \in S_\alpha$.  
Inserting complete sets of states in the appropriate sector 
of the theory (namely, that transforming as a pair of colour charges
in the fundamental and complex conjugate representation located at a
distance $R_E$), we find
\begin{equation}
  \label{eq:clust4}
  \begin{aligned}
&   \la 0 |\T_1 \left\{
    \hat{\W}_{E*}^{\,(T)}(r_\para,\vec r_\perp) 
\hat\Op_{S_\alpha}(X_{E\,{S}_\alpha})  
  \right\}| 0 \ra  \\
&\phantom{aua} = \sum_{s_1,s_2} \sum_{i,j} e^{-\f{T}{2}(E_{s_1}+E_{s_2})} \la
R_E ;{ij}| s_1\ra 
\la s_1|  
\T_1 \left\{
\hat\Op_{S_\alpha}(X_{E\,{S}_\alpha})  
 \right\} 
    | s_2\ra \la s_2
     | R_E ;{ij} \ra\,,  \end{aligned}
\end{equation}
and moreover
\begin{equation}
  \label{eq:clust4_bis}
   \la 0 |  \hat{\W}_{E*}^{\,(T)}(r_\para,\vec r_\perp) 
| 0 \ra = \sum_{s_1} \sum_{i,j} e^{-TE_{s_1}} \la
R_E ;{ij}| s_1\ra \la s_1  
     | R_E ;{ij} \ra\,,
\end{equation}
where $| R_E ;{ij} \ra\equiv [\hat{W}_E(R_E)]_{ij} |0 \ra$ is the
``flux-tube'' state created by the Wilson line $\hat{W}_E(R_E)$. 
In the limit $T\to\infty$, the dominant contribution comes from the
flux-tube ground state, $s_1=s_2=g=g(R_E)$ (since there is a gap with the
first excited state), and we obtain
\begin{equation}
  \label{eq:clust5}
 \C_E(X_{E\,S_\alpha};r_\para,\vec r_\perp) \equiv \lim_{T\to\infty}
 \C_E^{(T)}(X_{E\,S_\alpha};r_\para,\vec r_\perp)= \la g|  
\T_1 \left\{
\hat\Op_{S_\alpha}(X_{E\,{S}_\alpha})  
 \right\} 
    | g\ra \,.
\end{equation}
Consider now the case when the interpolating fields cluster into
subsets,  well separated from each other in the ``time''
direction. More precisely, given a partition ${\cal A}_K(S_\alpha)$ of 
$S_\alpha$ in $K$ parts, ${\cal
  A}_K(S_\alpha)=\{a_k\}_{k=1,\ldots,K}$, consider the limit
\begin{equation}
  \label{eq:sep2}
|x_{E\,\is\,1}-x_{E\,\isp\,1}|\to \infty\,, \quad \forall \is\in
a_k\,,~\forall \isp\in a_{k'}\,,~k\ne k'\,.
\end{equation}
By appropriately inserting complete sets of flux-tube states between
the subsets of interpolating fields, one can show that in this limit
the sums over intermediate states are dominated by the ground state,
and so 
\begin{equation}
  \label{eq:clust9}
  \begin{aligned}
\C_E(X_{E\,S_\alpha};r_\para,\vec r_\perp)
\to
\prod_{k=1}^K
\C_E(X_{E\,a_k};r_\para,\vec r_\perp) = \prod_{a\in {\cal A}_K(S_\alpha)}
\C_E(X_{E\,a};r_\para,\vec r_\perp)
\,.
\end{aligned}
\end{equation}
Let us now perform a decomposition in connected components in the usual
way, i.e., defining recursively, for any $T$, and for $A\subseteq S$,
\begin{equation}
  \label{eq:conn_dec}
   \C_E^{(T)\,{\rm conn}}(X_{E\,A};r_\para,\vec r_\perp) \equiv
\C_E^{(T)}(X_{E\,A};r_\para,\vec r_\perp) -
\sum_K\sum_{{\cal A}_K(A)\ne \{A\} }\prod_{a\in {\cal
    A}_K(A)}\C_E^{(T)\,{\rm conn}}(X_{E\,a};r_\para,\vec r_\perp) \,,
\end{equation}
where the sum is over all partitions of $A$, $\{A\}$ is the trivial
partition, and $\C_E^{(T)\,{\rm 
  conn}}=\C_E^{(T)}$ for one-element sets, so that
\begin{equation}
  \label{eq:conn_dec2}
  \C_E^{(T)}(X_{E\,S_\alpha};r_\para,\vec r_\perp) =\sum_K\sum_{{\cal
    A}_K(S_\alpha)}\prod_{a\in {\cal 
    A}_K(S_\alpha)}\C_E^{(T)\,{\rm conn}}(X_{E\,a};r_\para,\vec r_\perp) \,.
\end{equation}
In the limit $T\to\infty$ one has analogously
\begin{equation}
  \label{eq:conn_dec2Tinf}
  \C_E(X_{E\,S_\alpha};r_\para,\vec r_\perp) =\sum_K\sum_{{\cal
    A}_K(S_\alpha)}\prod_{a\in {\cal 
    A}_K(S_\alpha)}\C_E^{{\rm conn}}(X_{E\,a};r_\para,\vec r_\perp) \,.
\end{equation}
In this limit, $\C_E$ is translation-invariant along the ``time''
direction, i.e., direction 1, and so, by construction [see
Eq.~\eqref{eq:conn_dec}], each connected component $\C_E^{{\rm conn}}$
is also similarly invariant under ``time''-translations. Moreover,
Eq.~\eqref{eq:clust9} shows that in the limit $T\to\infty$, each
connected component vanishes when at least one of the interpolating
fields is very far from the others in the ``time'' direction. 
Let us make this discussion explicit by writing
\begin{equation}
  \label{eq:C_largeT}
\C_E^{(T)\,{\rm conn}}(X_{E\,a}) = C_T(t_{a},\hat{X}_{a})\,,  
\end{equation}
where $ t_{a} = \f{1}{N_a}\sum_{\is\in a} x_{E\,\is\,1} $ 
is the average ``time''-coordinate of the particles in part $a$, with
$N_a$ the corresponding number of particles, 
and $\hat{X}_{a}$ denotes collectively all the remaining
components of the coordinates. Here we have dropped the dependence on the
dipole size for simplicity. As $T\to\infty$,
\begin{equation}
  \label{eq:C_largeT2}
\lim_{T\to\infty} C_T(t_{a}, \hat{X}_{a}) = C(\hat{X}_{a})\,.  
\end{equation}
We can also say something about how this limit is approached. At
finite $T$, $C_T(t_{a}, \hat{X}_{a})$ is essentially constant
for $|t_{a}|\ll \f{T}{2}$, and should not change appreciably as long
as $|t_a| < \f{T}{2} - \kappa a_{\rm corr}$, where $a_{\rm corr}$ is
the so-called ``vacuum correlation
length''~\cite{DDSS,vaccorr1,vaccorr2} and $\kappa$ is some number of 
order 1, that 
depends also on the spread of the ``temporal'' components of the
positions of the interpolating fields (which again can be at most a few
$a_{\rm corr}$ since we are considering a connected correlation
function), but that is independent of $T$ (when $T$ is large enough
and only one short edge at a time is relevant to this issue). After a
transient region of size approximately $2\kappa a_{\rm corr}$, the
correlator drops essentially to zero when $|t_a| > \f{T}{2} + \kappa
a_{\rm corr}$. The conclusion is that $C_T(T \tau_{a}, \hat{X}_{a})$
tends to a constant function over the interval $\tau_a \in
[-\f{1}{2},\f{1}{2}]$: the transient regions in 
terms of $\tau_{a}$ shrink as $T\to \infty$, and the slope of the
function there diverges. So $C_T(t_{a}, \hat{X}_{a}) \to
\chi(\f{t_a}{T})C(\hat{X}_{a})$, or more precisely
\begin{equation}
  \label{eq:C_largeT3}
\lim_{T\to\infty} C_T(T\tau_{a}, \hat{X}_{a}) = 
\chi(\tau_a)C(\hat{X}_{a})\,,  
\end{equation}
with $\chi(\tau_a)$ the characteristic function of the interval
$[-\f{1}{2},\f{1}{2}]$. 

Cluster decompositions for $L_{E}^{\,(T)}$ and $M^{\,(T)}$ can also be
written down, in full analogy with Eqs.~\eqref{eq:conn_dec}
and \eqref{eq:conn_dec2}. Comparing them with the cluster
decomposition of $\C_E^{(T)}$, one finds 
\begin{equation}
  \label{eq:clust_M_2}
  \begin{aligned}
&  L_{E}^{\,(T)\,{\rm conn}}(P_{E \,a }; r_\para,\vec r_\perp) = 
\int dX_{E\,a}\, e^{iP_{E\,a}\cdot X_{E\,a}}\,
\C_E^{\,(T)\,{\rm conn}}(X_{E\,a};r_\para,\vec r_\perp)\,, \\
& M^{\,(T)\,{\rm conn}}(\Omega_{a};r_\para,\vec r_{\perp})= 
{\rm Lim}_{a}
\prod_{\is\in a} [-i\pi^{(s)}(\vec
p_{\is},s_{3\is})]
L_E^{\,(T)\,{\rm conn}}(e^{-i\f{\pi}{2}}(-P^0_{a}),\vec P_{a};r_\para,\vec r_{\perp})\,.
  \end{aligned}
\end{equation}
Here we have made use of the fact that the on-shell projector is factorised.
The connected components of $\bar{M}^{\,(T)}$ are easily obtained
using Eq.~\eqref{eq:wilson_bar_lsz3},
\begin{equation}
  \label{eq:clust_M_4}  
  \begin{aligned}
    \bar M^{\,(T)\,{\rm conn}}(\Omega_{{a}};r_\para,\vec r_{\perp})&\equiv
 \left[M^{\,(T)\,{\rm conn}}(\Omega_{{a}};r_\para,-\vec r_{\perp})\right]^*
\,.
  \end{aligned}
\end{equation}
Finally, a similar decomposition can be carried out for the various
quantities in the limit $T\to\infty$.

The ``time''-translation invariance of $\C_E^{{\rm conn}}$, for a
certain part, $a$, in some partition, ${\cal A}_K$, reflects itself in
the appearance of delta functions in $L_{E}^{\rm
  conn}\equiv\lim_{T\to\infty} L_{E}^{\,(T)\,{\rm conn}}$, imposing
the vanishing of the total ``temporal'' momentum of the particles in
$a$. Furthermore, as $\C_E^{{\rm conn}}$ vanishes when the ``time''
separation between the interpolating fields becomes large [see 
Eq.~\eqref{eq:clust9}], the corresponding integration regions give no
contribution to $L_{E}^{\rm conn}$, and no further delta functions of
subsets of ``temporal'' momenta can appear. Finally, as the 
analytic continuation required to obtain the matrix elements
$M=\lim_{T\to\infty}M^{\,(T)}$ does not involve $p_{E\,\is\,1}$, these
properties are inherited by the connected components $M^{\rm
  conn}\equiv\lim_{T\to\infty} M^{\,(T)\,{\rm conn}}$, which contain
one and the same delta function of the ``temporal'' momenta as
$L_{E}^{{\rm conn}}$. 
More precisely, for the physically relevant quantity 
$M^{\,(T)\,{\rm conn}}$ one can write
\begin{equation}
\label{eq:M_FT}
M^{\,(T)\,{\rm conn}}(\Omega_{a};r_{1\para},\vec r_{1\perp})=
  \int dt_a\,e^{iq_a t_a } F_T(t_a,P_{a}; r_{1\para},\vec r_{1\perp})\,,
\end{equation}
where $q_a\equiv {\textstyle\sum}_{\is\in 
      a}\,p_{\is\,1}$, 
for a certain function $F_T$, obtained from $C_T$ through integration
over $\hat X_a$, Wick-rotation of the momenta, and on-shell projection
(see Appendix \ref{app:largeT} for more details). The important point
is that these steps should not change the way the
large-$T$ limit is approached, i.e., for large $T$
\begin{equation}
  \label{eq:largeT_phys}
  \begin{aligned}
&\lim_{T\to\infty}  F_T(T\tau_a,P_{a}; r_{1\para},\vec r_{1\perp})=
\chi\left(\tau_a\right){\cal M}^{\,{\rm conn}}(\Omega_{a};r_{1\para},\vec r_{1\perp})
\,,    
  \end{aligned}
\end{equation}
for a certain ${\cal M}^{\,{\rm conn}}$, from which it follows
\begin{equation}
  \label{eq:mtil}
  \begin{aligned}
  \lim_{T\to\infty}  M^{\,(T)\,{\rm
      conn}}(\Omega_{a};r_{1\para},\vec r_{1\perp})&=
   \delta\left({\textstyle\sum}_{\is\in       a}\,p_{\is\,1}\right) 
{\cal M}^{\,{\rm conn}}(\Omega_{a};r_{1\para},\vec r_{1\perp})\,. 
  \end{aligned}
\end{equation}
For the other connected matrix element, $\bar M^{\,(T)\,{\rm conn}}$
[see Eq.~\eqref{eq:clust_M_4}], we similarly have 
\begin{equation}
  \label{eq:barcalM}
  \begin{aligned}
  \lim_{T\to\infty} \bar{M}^{\,(T)\,{\rm
      conn}}(\Omega_{{a}};r_{2\para},\vec r_{2\perp})&=
  \delta\left({\textstyle\sum}_{\is\in   {a}}\,p_{\is\,1}\right) 
\bar{\cal M}^{\,{\rm conn}}(\Omega_{{a}};r_{2\para},\vec r_{2\perp})
\\ \bar{\cal M}^{\,{\rm conn}}(\Omega_{{ a}};r_{2\para},\vec r_{2\perp}) &=
\left[{\cal M}^{\,{\rm 
      conn}}(\Omega_{{a}};r_{2\para},-\vec r_{2\perp})\right]^*
\,.
  \end{aligned}
\end{equation}

\subsection{Dipole-dipole potential from the Wilson-loop correlator}
\label{sec:Tdep}

The purpose of the analysis of the previous Subsection is twofold. On
the one hand, the cluster decomposition allows us to write down
explicitly the exponential form of the Wilson-loop correlator,
Eq.~\eqref{eq:ddpot3}. On the other hand, the properties of the
connected components in the large-$T$ limit imply that the correct
$T$-dependence is obtained.  

Let us start from the exponentiation. The decomposition of the matrix
elements into connected components is not yet the full story, since
what appears in Eq.~\eqref{eq:ddpot3} is the product of the matrix
elements $M^{\,(T)}$ and $\bar M^{\,(T)}$. Substituting 
the cluster decompositions of $M^{\,(T)}$ and $\bar M^{\,(T)}$
in Eq.~\eqref{eq:ddpot3}, one thus obtains a double sum over
partitions. Each pair of partitions ${\cal A}_K(S_\alpha)$,
$\bar{\cal A}_{\bar K}(S_\alpha)$ of $S_\alpha$, with $K$ and $\bar K$
parts, respectively, i.e., ${\cal
  A}_K(S_\alpha)=\{a_k\}_{k=1,\ldots,K}$ and $\bar{\cal A}_{\bar
  K}(S_\alpha)=\{\bar a_{\bar k}\}_{\bar k=1,\ldots,\bar K}$, 
can be uniquely rewritten as a partition ${\cal F}_J(S_\alpha)$
of $S_\alpha$ with $J$ parts, and a set of {\it irreducible} pairs of
partitions  $[{\cal A}_{K_j},\bar{\cal A}_{\bar{K}_j}](F_j)$
of the parts $F_j\in {\cal F}_J(S_\alpha)$. By an irreducible pair of
partitions we mean that there are no proper subpartitions 
$\{a'_k\}_{k=1,\ldots,K_j'}\subset {\cal A}_{K_j}(F_j)$, 
and $\{\bar{a}'_{\bar k}\}_{\bar{k}=1,\ldots,\bar{K}_j'}\subset \bar{\cal
  A}_{\bar{K}_j}(F_j)$, such that $\cup_k a'_k = \cup_{\bar k}
\bar{a}'_{\bar k}$. Checking a few examples should convince the
reader; a formal proof is given in Appendix
\ref{app:partitions}.  
The double sum over partitions
can therefore be rewritten as
\begin{equation}
  \label{eq:sum_pair_part_main}
  \sum_{K}\sum_{{\cal A}_K(S_\alpha)}\sum_{\bar K}\sum_{\bar{\cal
      A}_{\bar K}(S_\alpha)}=
  \sum_{J} \sum_{{\cal F}_J(S_\alpha)} \prod_{F\in{\cal F}_J(S_\alpha)}
\left(\sum_{K}\sum_{L}\sum_{
[{\cal A}_{K},\bar{\cal A}_{\bar K}](F)
}\right)\,.
\end{equation}
Working out the consequences of this fact is a straightforward but lengthy
exercise in combinatorics, which is described in detail in Appendix
\ref{app:expo}. Here we report only the final result for the
Wilson-loop correlator, which reads
\begin{equation}
  \label{eq:expo_final}
  \begin{aligned}
  {\cal G}^{(T)}(b;r_{1\para},\vec{r}_{1\perp},r_{2\para},\vec{r}_{2\perp})
&= \exp \left\{ \sum_{\alpha\ne \emptyset}\f{1}{\prod_s N_{s}(\alpha)!}
{\cal Q}_\alpha^{(T)}(b;r_{1\para},\vec{r}_{1\perp},r_{2\para},\vec{r}_{2\perp}) 
\right\} \,, 
  \end{aligned}
\end{equation}
where we have introduced the following quantities,
\begin{equation}
  \label{eq:expo_final_bis}
  \begin{aligned}
{\cal Q}_\alpha^{(T)}(b;r_{1\para},\vec{r}_{1\perp},r_{2\para},\vec{r}_{2\perp}) &=
\sum_{K}\sum_{\bar K}\sum_{[{\cal A}_{K},\bar{\cal A}_{\bar
    K}](S_\alpha)} 
\lella
\prod_{a\in {\cal     A}_K(S_\alpha)}
M^{\,(T)\,{\rm conn}}(\Omega_{a};r_{1\para},\vec r_{1\perp})
 \right.\right. \\ & 
\phantom{dioserpentone}
\left.\left.
\times\prod_{\bar a\in \bar{\cal A}_{\bar K}(S_\alpha)}
 \bar M^{\,(T)\,{\rm conn}}(\Omega_{\bar a};r_{2\para },\vec r_{2\perp})
\rirra_{\Omega_{S_\alpha};\,b}\,.
  \end{aligned}
\end{equation}
Recalling Eq.~\eqref{eq:ddpot2_0}, the dipole-dipole potential reads 
\begin{equation}
   \label{eq:pot_final}
  \begin{aligned}
        V_{dd}(&b;r_{1\para},\vec{r}_{1\perp},r_{2\para},\vec{r}_{2\perp})
= -\lim_{T\to\infty} \f{1}{T}\sum_{\alpha\ne \emptyset}\f{1}{\prod_s N_{s}(\alpha)!}
{\cal
  Q}_\alpha^{(T)}(b;r_{1\para},\vec{r}_{1\perp},r_{2\para},\vec{r}_{2\perp})
\\
&=\sum_{\alpha\ne \emptyset}\f{1}{\prod_s
  N_{s}(\alpha)!}\sum_{K}\sum_{\bar K}\sum_{[{\cal A}_{K},\bar{\cal 
    A}_{\bar K}](S_\alpha)}
V_{[{\cal A}_{K},\bar{\cal
    A}_{\bar K}](S_\alpha)}(b;r_{1\para},\vec r_{1\perp},r_{2\para},\vec r_{2\perp})\,,
  \end{aligned}
\end{equation}
where $[{\cal A}_K,\bar{\cal A}_{\bar K}](S_\alpha)$ is a pair of
irreducible partitions of $S_\alpha$, and 
\begin{equation}
  \label{eq:pot_contrib}
  \begin{aligned}
    &-V_{[{\cal A}_{K},\bar{\cal A}_{\bar
        K}](S_\alpha)}(b;r_{1\para},\vec r_{1\perp},r_{2\para},\vec r_{2\perp}) 
  \\ & \equiv  \lim_{T\to\infty}\f{1}{T}\lella 
\prod_{a\in {\cal     A}_K(S_\alpha)}
M^{\,(T)\,{\rm conn}}(\Omega_{a};r_{1\para},\vec r_{1\perp})
\prod_{\bar a\in \bar{\cal A}_{\bar K}(S_\alpha)}
 \bar M^{\,(T)\,{\rm conn}}(\Omega_{\bar{a}};r_{2\para},\vec r_{2\perp})
\rirra_{\Omega_{S_\alpha};\,b}\,. 
  \end{aligned}
\end{equation}
The crucial point is now to show that ${\cal Q}_\alpha^{(T)}$ diverges
linearly with $T$. As we have argued in the previous Subsection, in
the large-$T$ limit each connected component develops a Dirac delta of
the total ``temporal'' momenta $q_a \equiv {\textstyle\sum}_{\is\in   {
    a}}\,p_{\is\,1}$ and $\bar{q}_{\bar a}\equiv
{\textstyle\sum}_{\is\in   {\bar a}}\,p_{\is\,1}$ in each part.  
In Appendix \ref{app:partitions} we show that, due to the
irreducibility of the pair of partitions, only $K+\bar K-1\le {\cal
  N}_\alpha$ out of the $K+\bar{K}$ linear combinations of momenta 
$q_a$ and $\bar{q}_{\bar a}$ are independent, the only relation of 
linear dependence being
\begin{equation}
  \label{eq:lindeprel}
\sum_{a\in {\cal A}_K(S_\alpha)} q_a =
\sum_{\bar a\in \bar{\cal A}_{\bar K}(S_\alpha)} \bar q_{\bar a} =
\sum_{\is\in S_\alpha} p_{\is\,1}\,.
\end{equation}
In practical terms, this means that in the large-$T$ limit the
integral in Eq.~\eqref{eq:pot_contrib} is divergent, as one of the
$K+\bar{K}$ Dirac deltas of Eqs.~\eqref{eq:mtil} and \eqref{eq:barcalM}
has to be evaluated at zero. However, this also means that the
divergence is linear in $T$, so that it gets cancelled by the $1/T$
factor, and $V_{[{\cal A}_{K},\bar{\cal A}_{\bar K}](S_\alpha)}$ is
finite. A detailed calculation showing this, which makes use of the 
large-$T$ behaviour of the connected matrix elements,
Eq.~\eqref{eq:largeT_phys}, is reported in Appendix
\ref{app:largeT}. Here we quote only the final result,  
\begin{equation}
 \label{eq:delta_fin}
  \begin{aligned}
&-V_{[{\cal A}_{K},\bar{\cal A}_{\bar K}](S_\alpha)}
(b;r_{1\para},\vec r_{1\perp},r_{2\para},\vec r_{2\perp})  \\ &=   
\int d\Omega_{S_\alpha}\,e^{-bE(\Omega_{S_\alpha})}\,
(2\pi)^{K+\bar{K}-1}
 \delta_{[{\cal A}_{K},\bar{\cal A}_{\bar K}](S_\alpha)}(p_1)
\\ 
&\phantom{uuuuuuuuuuuuuuu}\times
\prod_{a\in {\cal A}_K(S_\alpha)} {\cal M}^{\,{\rm
      conn}}(\Omega_{a};r_{1\para},\vec r_{1\perp}) 
\prod_{\bar a\in \bar{\cal A}_{\bar K}(S_\alpha)}
\bar{\cal M}^{\,{\rm conn}}(\Omega_{\bar a};r_{2\para},\vec r_{2\perp})
\,,
\\ &\delta_{[{\cal A}_{K},\bar{\cal A}_{\bar K}](S_\alpha)}(p_1)
\equiv 
\delta\left({\textstyle\sum}_{\is\in
      S_\alpha}p_{\is\,1}\right)
\prod^\circ_{a\in {\cal A}_K(S_\alpha)}
\delta
\left({\textstyle\sum}_{\is\in
      a}p_{\is\,1}\right)
\prod^\circ_{\bar a\in \bar{\cal A}_{\bar K}(S_\alpha)}
\delta
\left({\textstyle\sum}_{\is\in
      \bar a}p_{\is\,1}\right)
 \,,
  \end{aligned}
\end{equation}
where the symbol $\circ$ denotes that the product is over all the
parts in the partition {\it but one}. 

The expressions Eqs.~\eqref{eq:pot_final} and \eqref{eq:delta_fin}
fully encode the static dipole-dipole potential when the dipoles do
not overlap in the direction of their separation, i.e., for all $\vec b$
and $\vec{r}_{1,2}$ such that $|\vec b\,|>
|\vec r_{1}\cdot \hat b|+|\vec r_{2}\cdot \hat b|$. In the next
Section we use them to extract the behaviour of the
potential at asymptotically large distances.

\section{Asymptotic behaviour of the potential at large distance}
\label{sec:largeb}

At this point it is straightforward to derive the large-$b$ behaviour
of the potential. From Eq.~\eqref{eq:delta_fin} we see that 
the $b$-dependence is contained entirely in the factor
$e^{-bE(\Omega_{S_\alpha})}$. However, we still have to perform the
phase-space integration. Making the change of variables
$\sqrt{b}\vec p_{\is}= \vec q_{\is}$, we can rewrite
Eq.~\eqref{eq:delta_fin} as follows, 
\begin{equation}
\label{eq:ene_largeb_4}
  \begin{aligned}
-V_{[{\cal A}_{K},\bar{\cal A}_{\bar
    K}](S_\alpha)}&(b;r_{1\para},\vec r_{1\perp},r_{2\para},\vec r_{2\perp}) 
\\ 
=&~
   b^{-\f{3{\cal N}_\alpha  - (K+\bar K-1) }{2}}
 \int
    d\hat\Omega_{S_\alpha} 
\,e^{-b
\sum_{\is\in S_\alpha} m_{(s)}\hat\varepsilon_{\is}
}\,
(2\pi)^{K+\bar K-1}
\delta_{[{\cal A}_{K},\bar{\cal A}_{\bar K}](S_\alpha)}(q_1)
\\ 
&\times
\prod_{a\in {\cal A}_K(S_\alpha)}
{\cal M}^{\,{\rm
    conn}}\left(
\tf{1}{\sqrt{b}}\hat\Omega_a
;r_{1\para},\vec r_{1\perp}
\right)   
\prod_{\bar a\in \bar{\cal A}_{\bar K}(S_\alpha)}
\bar{\cal M}^{\,{\rm
    conn}}\left(
\tf{1}{\sqrt{b}}\hat\Omega_{\bar a}  
;r_{2\para},\vec r_{2\perp}
\right) 
\,,
  \end{aligned}
\end{equation}
where 
\begin{equation}
  \label{eq:ene_largeb_2}
  \begin{aligned}
\int d\hat\Omega_{S_\alpha} &\equiv  
\sum_{\{s_3\}}  \prod_{\is\in S_\alpha}
\int \f{d^3q_{\is}}{(2\pi)^32m_{(s)}\hat\varepsilon_{\is}}
\,,&&&
\hat\varepsilon_{\is} &\equiv \sqrt{1 + \f{\vec
      q_{\is}^{\,2}}{b m_{(s)}^2} }\,,
  \end{aligned}
\end{equation}
$\sum_{\{s_3\}}$ denotes the sum over the spins of all
particles, and we have denoted
\begin{equation}
  \label{eq:omegaz}
   \tf{1}{\sqrt{b}}\hat\Omega_{A}  =
\{(\tf{1}{\sqrt{b}}\vec q_{\is},s_{3\is}) ~|~ \is\in A\}\,,\quad
A\subseteq S\,.
\end{equation}
In the limit of large $b$, we can expand $\hat\varepsilon_{\is}$ and
the integration measure $d\hat\Omega_{S_\alpha}$ as follows,
\begin{equation}
  \label{eq:ene_largeb_5}
  \begin{aligned}
    \hat\varepsilon_{\is} & = 1 + \f{\vec
      q_{\is}^{\,2}}{2bm^2_{(s)}} + \Op(b^{-2}) \,,\\
d\hat\Omega_{S_\alpha} &= \prod_{\is\in
      S_\alpha} \f{d^3q_{\is}}{(2\pi)^32m_{(s)}(1+\Op(b^{-1}))} \equiv
dq_{S_\alpha}\left(1+\Op(b^{-1})\right)
\,,
  \end{aligned}
\end{equation}
and moreover we can expand the matrix elements around zero momentum,
\begin{equation}
\label{eq:ene_largeb_5_bis}
\tf{1}{\sqrt{b}}\hat\Omega_A 
=
\{(0,s_{3\is}) ~|~ \is\in A\} + \Op(b^{-\f{1}{2}})
\equiv \Omega^0_{A}+ \Op(b^{-\f{1}{2}})\,.
\end{equation}
To leading order we find
\begin{equation}
\label{eq:ene_largeb_8}
  \begin{aligned}
-V_{[{\cal A}_{K},\bar{\cal A}_{\bar K}](S_\alpha)}
(b;r_{1\para},\vec r_{1\perp},r_{2\para},\vec r_{2\perp})
  \mathop\to_{b\to\infty}
&~~     b^{-\f{3{\cal N}_\alpha  - (K+\bar K-1) }{2}} 
e^{-b \sum_{s}m_{(s)}N_s(\alpha)}
 \mathscr{M}_{[{\cal A}_{K},\bar{\cal A}_{\bar K}](S_\alpha)}
\\ &~~~\times 
    {\cal M}_{[{\cal A}_{K},\bar{\cal A}_{\bar K}](S_\alpha)}
(r_{1\para},\vec r_{1\perp},r_{2\para},\vec r_{2\perp}) 
\,,
  \end{aligned}
\end{equation}
where the full $b$-dependence is in the first two factors, 
$\mathscr{M}_{[{\cal A}_{K},\bar{\cal A}_{\bar K}](S_\alpha)}$ 
is a constant,
\begin{equation}
  \label{eq:mscr}
  \begin{aligned}
&    \mathscr{M}_{[{\cal A}_{K},\bar{\cal A}_{\bar K}](S_\alpha)}
\equiv
    \int
    dq_{S_\alpha}
    \,e^{-
      \sum_{\is\in S_\alpha}\f{\vec
      q_{\is}^{\,2}}{2m_{(s)}}}\,
    (2\pi)^{K+\bar K-1}
\delta_{[{\cal A}_{K},\bar{\cal A}_{\bar K}](S_\alpha)}(q_1) \\
&\phantom{uuuuu}= 
\f{1}{2^{{\cal N}_\alpha} (2\pi)^{2{\cal N}_\alpha-(K+\bar K-1)}} 
 \int  \left[\prod_{\is\in
      S_\alpha} dq_{\is 1}\, e^{-
\f{q_{\is 1}^{\,2}}{2m_{(s)}}}\,\right]  
\delta_{[{\cal A}_{K},\bar{\cal A}_{\bar K}](S_\alpha)}(q_1)
\,,
  \end{aligned}
\end{equation}
and the dependence on the size and orientation of the dipoles is
contained in ${\cal M}_{[{\cal A}_{K},\bar{\cal A}_{\bar
    K}](S_\alpha)}$, 
\begin{equation}
  \label{eq:mcal}
  \begin{aligned}
    {\cal M}_{[{\cal A}_{K},\bar{\cal A}_{\bar K}](S_\alpha)}&
(r_{1\para},\vec r_{1\perp},r_{2\para},\vec r_{2\perp}) \\ &
  \equiv 
  \sum_{\{s_3\}} 
\prod_{a\in {\cal A}_K(S_\alpha)}
    {\cal M}^{\,{\rm
        conn}}\left(\Omega^0_{a};r_{1\para},\vec r_{1\perp}
    \right)   
\prod_{\bar a\in \bar{\cal A}_{\bar K}(S_\alpha)}
    \bar{\cal M}^{\,{\rm
        conn}}\left(\Omega^0_{\bar a};r_{2\para},\vec r_{2\perp}
    \right) 
\,.
  \end{aligned}
\end{equation}
Here we are implicitly assuming that the connected matrix elements
${\cal M}^{\,{\rm conn}}$ are finite, nonzero quantities at zero
momentum. This is expected to be the case for states containing only
massive particles.\footnote{\label{foot:massless} In the presence of
  massless particles, they are expected to vanish, in order to cancel
  the divergence in the phase-space measure. We have verified this
  explicitly in the simple case of pure $U(1)$ gauge theory, i.e., for
  free photons. Notice that, in the case discussed in the present
  paper, the above-mentioned divergence is only apparent and does not
  require the vanishing of the matrix elements. However, one can
  easily show that in the case of Wilson loops at nonzero angle
  $\theta$, considered, e.g., in Ref.~\cite{sigtot}, there is indeed a
  logarithmic divergence unless the matrix elements vanish. 
} 
Notice that the exponent $\gamma=[3{\cal N}_\alpha - (K+\bar K -1)]/2
$ of the power-law term in Eqs.~\eqref{eq:ene_largeb_4} and
\eqref{eq:ene_largeb_8} obeys the inequality $\gamma \ge {\cal 
  N}_\alpha \ge 1$ (see the end of Appendix \ref{app:partitions}),
as well as $\gamma\le (3{\cal N}_\alpha - 1)/2$ since $K,\bar K\ge 1$,
for any (non-vacuum) state.    

The leading behaviour of the potential is determined by the
contributions $V_{[{\cal A}_{K},\bar{\cal A}_{\bar K}](S_\alpha)}$ of 
the lightest states with nonzero Wilson-loop matrix elements, with
higher-order contributions being exponentially suppressed. 
Since the Wilson-loop operator depends only on the gauge fields, it is
obviously invariant under any symmetry of the theory acting only on
the matter degrees of freedom. This implies a selection rule involving
the corresponding quantum numbers, which have to be the same as those
of the vacuum in order for the Wilson-loop matrix element to be
nonzero. In particular, in the case of QCD the Wilson loop is
insensitive to flavour, and so its matrix elements can be nonzero only
for states carrying no flavour quantum numbers, which results in a
selection rule for baryon number, electric charge, strangeness, etc.,
that must all vanish.   

For the interesting gauge theories, the lightest particle is typically
a spin-zero particle. Indeed, lattice results for $SU(N_c)$ pure-gauge
theory indicate that the lightest ``glueball'' has quantum numbers
$J^{PC}=0^{++}$ (see, e.g., Ref.~\cite{LRR}). For theories with $N_f$
light fermions, the lightest 
particles are the $N_f^2-1$ (pseudo)Goldstone bosons generated by the
spontaneous breaking of the (approximate) chiral $SU(N_f)_L\times
SU(N_f)_R$ symmetry (at least if $N_f$ is not too large). This is the case for
real-world QCD ($N_c=3$, $N_f=2$), where the lightest states are the pions
(pseudoscalars). For spin-zero particles it is possible to derive
easily further selection rules on parity and charge conjugation. As we
show in Appendix \ref{sec:spzero}, for a self-conjugate
particle with $C$ and $P$ phases $\eta_{C}$ and $\eta_P$, nonzero
matrix elements are possible only if $\eta_C = \eta_P = 1$.

Let us now discuss in detail a few interesting cases. In QCD, the
lightest particles are the three pions, $\pi^0$ and $\pi^\pm$, but due to
the selection rules on electric charge and on parity, they
have vanishing one-particle matrix elements. 
The lightest state with nonzero matrix element is the one containing
two $\pi^0$, followed by the state containing a $\pi^+\pi^-$ pair.
In both cases there is a single irreducible pair of partitions
contributing to the potential, namely the pair of trivial partitions $[{\cal
  A}_1,\bar{\cal A}_1]$, and so denoting with $S_{\pi^0\pi^0}$ and
$S_{\pi^+\pi^-}$ the relevant $S_\alpha$, we find from
Eq.~\eqref{eq:mscr} 
\begin{equation}
  \label{eq:ene_largeb_9}
  \begin{aligned}
&\mathscr{M}_{[{\cal A}_1,\bar{\cal A}_1](S_{\pi^0\pi^0})} 
= 
\f{ \sqrt{m_{\pi^0}}}{(4\pi)^{\f{5}{2}}}
\,,
&&&
&
\mathscr{M}_{[{\cal A}_1,\bar{\cal A}_1](S_{\pi^+\pi^-})} 
= 
\f{ \sqrt{m_{\pi^\pm}}}{(4\pi)^{\f{5}{2}}}
\,.
  \end{aligned}
\end{equation}
Due to the very small relative mass difference between the neutral and
the charged pions (also when electromagnetic effects are neglected),
the $\pi^+\pi^-$ contribution is appreciably suppressed compared to
the $\pi^0\pi^0$ contribution only for distances well beyond the range
of the dipole-dipole interaction. Therefore, although strictly
speaking it is the $\pi^0\pi^0$ state that determines the
asymptotic behaviour of the potential, it is physically more
meaningful to treat charged and neutral pions on the same footing.
We will then consider the limit of exact isospin symmetry, and ignore
the small mass difference between $\pi^0$ and $\pi^\pm$.  
In this limit the contributions of the $\pi^0\pi^0$ state and of the
$\pi^+\pi^-$ state are identical,\footnote{As the Wilson loop is
  flavour-blind, the only contributions to the Wilson-loop matrix elements
  $M^{\,(T)}$ come from the isosinglet components of these states, which
  differ only by a sign.} and so, taking into account the
symmetry factor $1/2$ for the $\pi^0\pi^0$ state, we have to leading
order\footnote{According to the discussion above, in real QCD the strict
  asymptotic behaviour is obtained from
  Eq.~\eqref{eq:ene_largeb_final} by replacing the factor $3/2$ with
  $1/2$, and using the $\pi^0$ mass and $\pi^0\pi^0$ matrix elements.} 
\begin{equation}
  \label{eq:ene_largeb_final}
  \begin{aligned}
     V_{dd}(b;&r_{1\para},\vec{r}_{1\perp},r_{2\para},\vec{r}_{2\perp}) 
\mathop\to_{b\to\infty}  
V_{[{\cal A}_{1},\bar{\cal
     A}_{1}](S_{\pi\pi})}(b;r_{1\para},\vec r_{1\perp},r_{2\para},\vec r_{2\perp})
\\ & \mathop\to_{b\to\infty} 
-\f{3}{2}\sqrt{m_\pi}\f{e^{-2m_\pi b} }{(4\pi b)^{\f{5}{2}}} 
{\cal M}^{\,{\rm
    conn}}\left(\Omega^0_{S_{\pi\pi}};r_{1\para},\vec r_{1\perp}
\right)   
\bar{\cal M}^{\,{\rm
    conn}}\left(\Omega^0_{S_{\pi\pi}};r_{2\para},\vec r_{2\perp}
\right)\,.
  \end{aligned}
\end{equation}
Due to Eq.~\eqref{eq:barcalM}, the potential is attractive at large
distances.\footnote{\label{foot:attr} More precisely, this is certainly
  true for small enough dipole sizes if the matrix elements are
  nonvanishing and analytic in $r_\para$ and $\vec r_\perp$ at
  zero. If the matrix 
  elements are continuous and never vanish, then this is true for all
  dipole sizes. 
  Furthermore, notice that Eq.~\eqref{eq:barcalM} 
  implies that this is
  true for $r_{2\para}=r_{1\para}$, $\vec r_{2\perp}=-\vec r_{1\perp}$. 
  As we show in Appendix \ref{sec:spzero},  
  ${\cal M}^{\,{\rm conn}}\left(\Omega^0_{S_{\pi\pi}};r_{\para},\vec
    r_{\perp} \right)$ depends on $\vec r_{\perp}$ only through $\vec
  r_{\perp}^{\,2}$, so this is again true for 
  $r_{2\para}=r_{1\para}$, 
  $|\vec r_{2\perp}|=|\vec r_{1\perp}|$.
 }
For $N_f$ degenerate flavours of quarks $q_i$, the relevant states are
those with pairs of ``pions'' $\pi_{ij}\pi_{ji}$, where
$\pi_{ij}=q_i\bar q_j$ for $i\ne j$, and $N_f-1$ pairs $\pi_i\pi_i$ with 
$\pi_i$ a combination of $q_i\bar q_i$ (the completely symmetric one
is excluded).  
There are
$N_f(N_f-1)/2$ pairs with $i\ne j$, and the $N_f-1$ states with two
$\pi_i$ require a symmetry factor $1/2$; the net effect is to replace
\begin{equation}
  \label{eq:replace}
\f{3}{2} \to \f{N_f(N_f-1)}{2} + \f{N_f-1}{2} = \f{N_f^2-1}{2}  
\end{equation}
in Eq.~\eqref{eq:ene_largeb_final}. 

The dependence on $b$ and the properties of 
our result, Eqs.~\eqref{eq:ene_largeb_final} and \eqref{eq:replace},
agree with the findings of Refs.~\cite{Pot4,FK}, which apply in the
regime of small dipole sizes. On the other hand, the calculations of
Refs.~\cite{LZ1,LZ2}, via AdS/QCD and in the ILM, respectively, report
large-distance behaviours of the form 
$V_{dd}^{\rm AdS/QCD} \sim {e^{-M_X b}}/{b^{\f{3}{2}}}$ 
[see Ref.~\cite{LZ1}, Eq.~(37)]  and
$V_{dd}^{\rm ILM} \sim {e^{-m_s b}}/{b^{\f{1}{2}}}$ 
[see Ref.~\cite{LZ2}, Eq.~(64)], where 
$M_X=m_{\rho}\sqrt{{17}/{8}}$ with 
$m_\rho$ the rho mass, and $m_s\sim 350\,{\rm MeV}$.\footnote{
\label{foot:powers} The
  fractional powers of $b$ look troublesome, since they cannot 
appear in our general formulas, Eqs.~\eqref{eq:ene_largeb_4} and
\eqref{eq:ene_largeb_8}. Indeed, turning around the inequalitites
reported after Eq.~\eqref{eq:mcal}, we have $(2\gamma +1)/3 \le {\cal
  N}_\alpha \le \gamma$, that cannot be satisfied by any integer ${\cal
  N}_\alpha$ for $\gamma=\f{1}{2},\f{3}{2}$. 
However, in our
  opinion 
these values are due to small mistakes in the extraction of the
asymptotic behaviour of the potential. Correcting these mistakes 
we find $V_{dd}^{\rm AdS/QCD} \sim {e^{-M_X b}}/{b}$, and
$V_{dd}^{\rm ILM} \sim {e^{-m_s b}}/{b}$, which match the form of 
one-particle contributions to the potential.}
Concerning the ILM result, the mass scale $m_s$ is approximately of
the right magnitude.\footnote{The mass $m_s$ corresponds to a scalar 
glueball state in the ILM~\cite{LZ2}, which is stable to leading order in
$1/N_c$, but which develops a nonvanishing decay width in higher
orders~\cite{KPZ}, that turns it into a resonant two-pion state: this
could explain the ``anomalous'' power-law correction $b^{-1}$ to the
exponential decay in the corrected expression for $V_{dd}^{\rm ILM}$
reported in footnote \ref{foot:powers}.} 
On the other hand, the mass scale $M_X$ is clearly much larger than
the pion threshold. However, the AdS/QCD correspondence is expected to
hold in the large-$N_c$ and strong-coupling limits, and so it is
perhaps more appropriate to compare this result to the one we have
obtained in pure-gauge theory, discussed below. In this case the mass
scale $M_X$ is of the right order of magnitude, although still quite
smaller than the lightest glueball mass at large $N_c$ (which is
slightly larger than at $N_c=3$~\cite{LRR}, see below). In the
estimate one should probably use the quenched value $m^{\rm q}_\rho$
for the rho mass, which, however, does not differ too much from the
physical value: using the quenched lattice results of
Ref.~\cite{GGLMP} for quenched pion masses $m^{\rm q}_\pi$ below
$m^{\rm q}_\pi\lesssim 400\,{\rm MeV}$, one has $m^{\rm q}_\rho\simeq
800\div 900\, {\rm MeV}$, resulting in $M_X\sim 1.2\div 1.3\,{\rm
  GeV}$, i.e., about $30\%$ below the lightest glueball mass. 

\begin{figure}[t]
  \centering
  \includegraphics[width=0.47\textwidth]{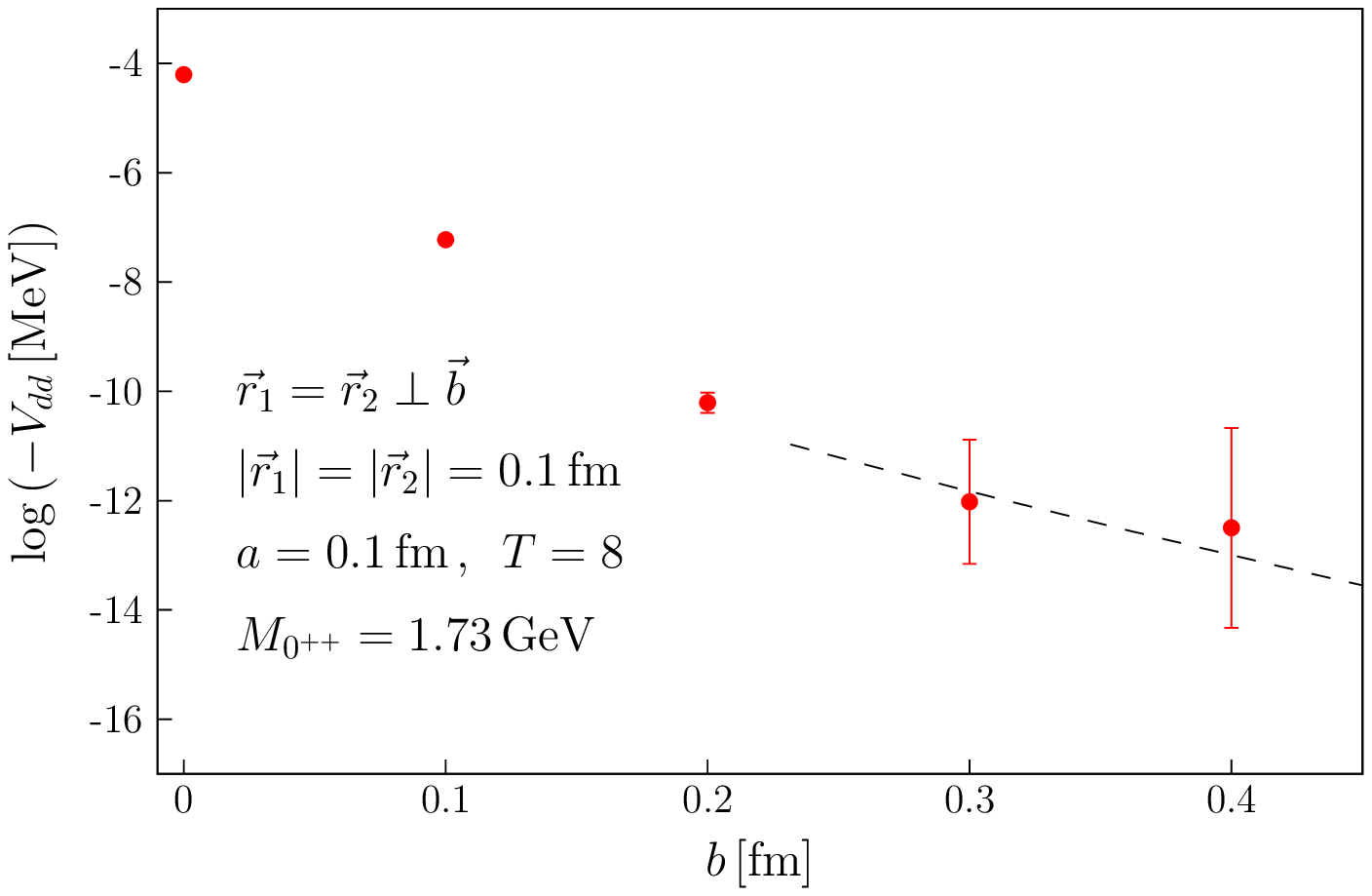}\hfil  \includegraphics[width=0.47\textwidth]{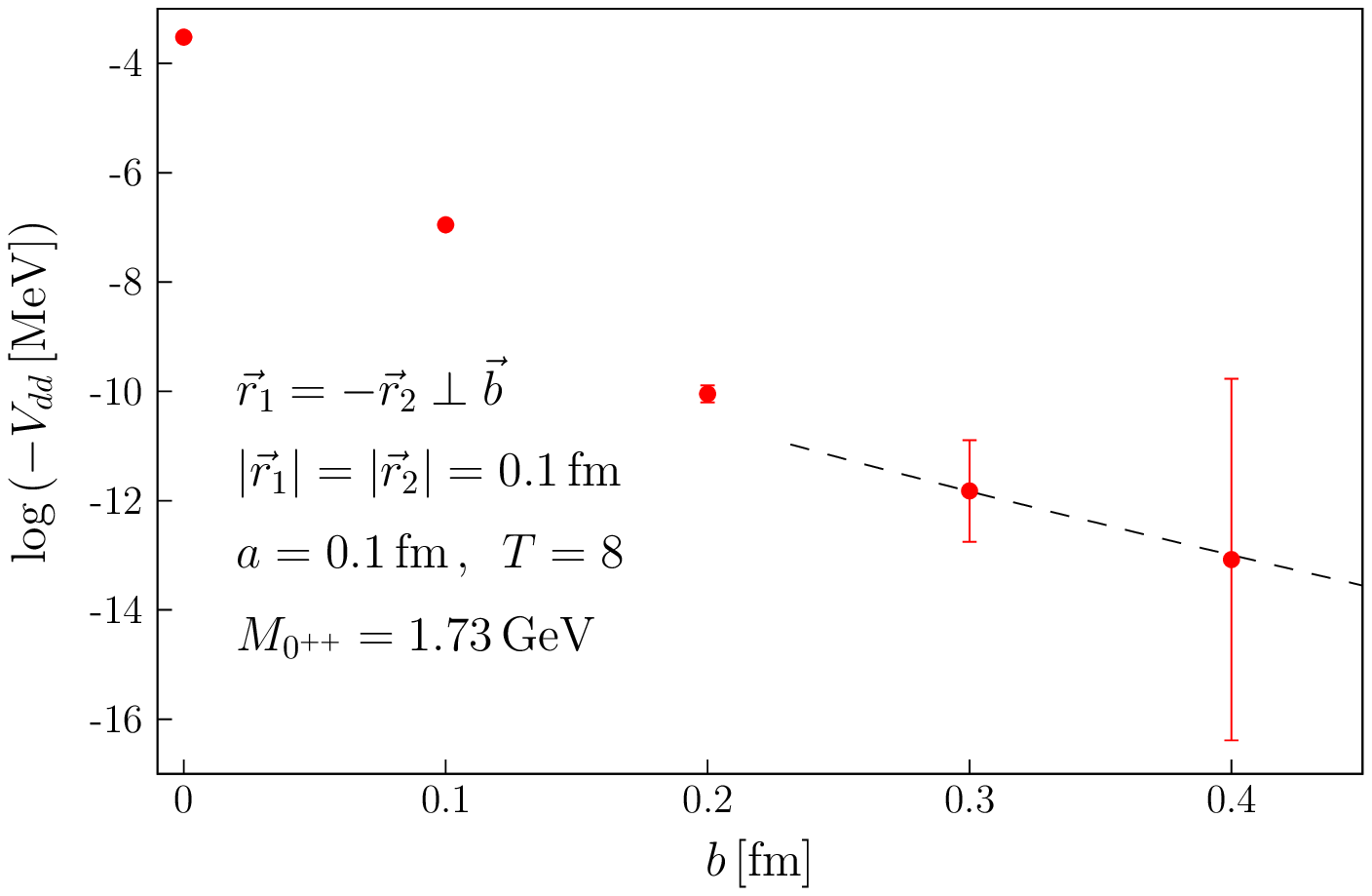} 
  \caption{Lattice determination of the static dipole-dipole potential
    in {\it quenched} QCD (data are taken from
    Ref.~\cite{GM2010}). Only statistical errors are shown. The dashed line 
    corresponds to the asymptotic behaviour,
    Eq.~\eqref{eq:ene_largeb_12}, with the numerical prefactor adjusted to
    fit the data points at $b=0.3,\,0.4~{\rm fm}$.}  
  \label{fig:2}
\end{figure}

In pure $SU(N_c)$ gauge theory, the lightest state contributing to the
potential is the one containing a single $0^{++}$ glueball, which we
denote by $S_{0^{++}}$. In this case there is obviously a single
relevant pair of partitions, and so
\begin{equation}
  \label{eq:ene_largeb_11}
  \begin{aligned}
&\mathscr{M}_{[{\cal A}_1,\bar{\cal A}_1](S_{0^{++}})} 
= \f{1}{4\pi}\,,
  \end{aligned}
\end{equation}
so that
\begin{equation}
  \label{eq:ene_largeb_12}
  \begin{aligned}
&
V_{dd}(b;r_{1\para},\vec{r}_{1\perp},r_{2\para},\vec{r}_{2\perp})
\mathop\to_{b\to\infty}  
V_{[{\cal A}_{1},\bar{\cal A}_{1}](S_{0^{++}})}(b;r_{1\para},\vec
r_{1\perp},r_{2\para},\vec r_{2\perp}) 
\\
&\phantom{uuuuuu}\mathop\to_{b\to\infty} -\f{ e^{- m_{0^{++}} b } }{4\pi b}
{\cal M}^{\,{\rm
    conn}}\left(\Omega^0_{S_{0^{++}}};r_{1\para},\vec r_{1\perp}
\right)   
\bar{\cal M}^{\,{\rm
    conn}}\left(\Omega^0_{S_{0^{++}}};r_{2\para},\vec r_{2\perp}
\right) \,.
  \end{aligned}
\end{equation}
Also in this case the potential is attractive.\footnote{See footnote
\ref{foot:attr}. In Appendix \ref{sec:spzero} we show that also ${\cal
  M}^{\,{\rm conn}}(\Omega^0_{S_{0^{++}}};r_{\para},\vec 
    r_{\perp} )$ depends on $\vec r_{\perp}$ only through $\vec
  r_{\perp}^{\,2}$.}
For $N_c=3$, i.e., in {\it quenched} QCD, the mass
of the lightest glueball is  
$m_{0^{++}}\simeq 1.73\,{\rm GeV}$~\cite{MP}, corresponding to an
interaction range $m_{0^{++}}^{-1}\simeq 0.11\,{\rm fm}$, so that the
asymptotic regime should be reached at distances accessible to lattice
calculations. In Fig.~\ref{fig:2} we compare the functional
dependence of Eq.~\eqref{eq:ene_largeb_12} with the numerical results
obtained on the lattice in Ref.~\cite{GM2010}. 
The potential is determined from Wilson loops of length $T=8$ and
width $|\vec r_{1,2}|=1$ in lattice units, on configurations obtained on a
$16^4$ lattice at $\beta=6.0$, corresponding to lattice spacing
$a\simeq 0.1\,{\rm fm}$. 
Lattice results and analytical prediction are compatible, 
although within rather large numerical errors. 

The most important subleading corrections come from the expansion in
inverse powers of $b$ of the energy, the phase-space measure and the
matrix elements, keeping fixed the particle content, i.e., for
two-pion states in QCD and for the lightest glueball state in
pure-gauge theory. From Eqs.~\eqref{eq:ene_largeb_5} and
\eqref{eq:ene_largeb_5_bis}, and since terms 
linear in the momenta in the expansion of the matrix elements give
vanishing contributions upon integration, we have that the first subleading
term is of relative order $b^{-1}$. 

From Eq.~\eqref{eq:ene_largeb_8} we see that for a given particle
content, with total number of particles ${\cal N}_\alpha$, the leading
(in $b$) contribution comes from the irreducible pair of partitions
with maximal $K+\bar K$,  
which cannot exceed ${\cal N}_\alpha+1$. In pure-gauge theory, where
states with a nonvanishing one-particle matrix 
element are present, the maximal value is attained, e.g., by the pair
of partitions where one is trivial (the whole set) and one is maximal
(each element is a part).
In QCD [and in similar theories with (pseudo-)Goldstone bosons] there
are no such 
states, and nonvanishing matrix elements are at least of the
two-particle type. As a consequence, one has $K,\bar K\le  [{\cal
  N}_\alpha/2]$, so that $K+\bar K\le {\cal N}_\alpha$ if   ${\cal
  N}_\alpha$ is even, and $K+\bar K\le {\cal N}_\alpha-1$ if   ${\cal
  N}_\alpha$ is odd.
The leading contribution at the ${\cal
  N}_\alpha$-particle level is thus proportional to
\begin{equation}
    \label{eq:ene_largeb_13}
    \begin{aligned}
\text{pure-gauge}:&~~~ \f{e^{- {\cal N}_\alpha m_{0^{++}}   b}}{b^{{\cal
      N}_\alpha}}\,, &&&
\text{QCD}:&~~~ \f{e^{- {\cal N}_\alpha m_{\pi}  b}}{b^{{\cal
      N}_\alpha+\f{1}{2}}}\bigg|_{\text{${\cal N}_\alpha$ even}}\,,
\quad \f{e^{- {\cal N}_\alpha m_{\pi}  b}}{b^{{\cal
      N}_\alpha+1}}\bigg|_{\text{${\cal N}_\alpha>1$, odd}}
\,.
    \end{aligned}
\end{equation}
It is worth discussing briefly what happens in the presence of
massless particles. In this case we expect the matrix elements to
vanish as powers of the momenta for small $|\vec p\,|$ (see footnote
\ref{foot:massless}). Here we drop the particle indices for
simplicity. For a multiparticle state containing only such massless 
particles, we expect by symmetry that each of them
contributes the same power, $\lambda$, of $|\vec p\,|$, to the
small-momentum behaviour of the matrix elements. Rather than
rescaling the momenta as in Eq.~\eqref{eq:ene_largeb_4}, we now more
conveniently set $b\vec p= \vec q$. For large $b$ we find that 
$V_{dd} \sim b^{-\gamma}$, with $\gamma=1+2\lambda$ if one-particle
matrix elements are nonzero, and $\gamma=3+4\lambda$ if matrix
elements are nonzero starting from the two-particle level. An explicit
calculation shows that $\lambda=1$ for free photons, resulting in the
well-known large-distance behaviour of the dipole-dipole electrostatic
potential.\footnote{Notice however that in this case our derivation of
  the cluster decomposition fails, since there is no gap in the
  spectrum of intermediate states.} If the same value is assumed for
massless pions in the chiral limit, then we find $\gamma=7$, in
agreement with Refs.~\cite{Pot4,FK}.

\section{Conclusions}
\label{sec:concl}

In this paper we have derived a general nonperturbative formula for
the asymptotic large-distance behaviour of the potential between two
static colourless dipoles, valid for a wide class of non-Abelian gauge
theories, and for any dipole size. Our result is based only on the
symmetries and on the nature of the spectrum of the relevant theories,
and is therefore a robust result. In particular, calculations
involving any kind of approximation have to compare successfully to
our predictions.

In the case of QCD, we have found the same dependence on the distance
as in the results of Refs.~\cite{Pot4,FK}, which are valid in the
regime of small dipole sizes. 
We have also compared our results to the recent nonperturbative
calculations of Refs.~\cite{LZ1,LZ2}, which make use of the AdS/QCD
approach and of the Instanton Liquid Model, respectively. In both
cases we find qualitative agreement with our results (apart from some
``anomalies'' which remain to be clarified). 
 
We have also discussed the case of pure $SU(N_c)$ gauge theory, for
which, to the best of our knowledge, there were so far no estimates,
and compared our prediction with the available lattice
results (for $N_c=3$)~\cite{GM2010}, finding agreement (within the
rather limited accuracy of the numerical data). 

We conclude by observing that the techniques developed in this paper
could be easily generalised to the case of the correlator of two Euclidean
Wilson loops forming a nonzero angle $\theta$, which is relevant to the
study of soft high-energy scattering and hadronic total cross sections
(see Ref.~\cite{sigtot} and references therein).

\section*{Acknowledgments}

MG is supported by the Hungarian Academy of Sciences under
``Lend\"ulet'' grant No. LP2011--011.

\appendix

\section{Decomposition of pairs of partitions in irreducible
  subpartitions }
\label{app:partitions}

Let $S$ be a finite discrete set. 
We call {\it irreducible} a pair of
partitions ${\cal A}_K(S) =\{a_k\}_{k=1,\ldots,K}$ and 
$\bar{\cal A}_{\bar K}(S)=\{\bar a_{\bar k}\}_{\bar k=1,\ldots,\bar
  K}$ of $S$, with $K$ and $\bar K$ 
parts, respectively, if there are no proper subsets ${\cal I}_S\subset
I_{\cal A}=\{1,\ldots,K\}$ and 
$\bar {\cal I}_S\subset I_{\bar{\cal A}}=\{1,\ldots,\bar K\}$ such
that $\cup_{k\in{\cal I}_S} a_k =\cup_{\bar k\in \bar{\cal I}_S} \bar
a_{\bar k}$. An irreducible pair of partitions of
$S$ will be denoted by $[{\cal A}_K,\bar{\cal A}_{\bar K}](S)$.
We prove now the following statement:
\begin{quote}\it
Any pair of partitions ${\cal A}_K(S)$ and $\bar{\cal A}_{\bar K}(S)$ 
of a set $S$ can be written uniquely as a pair 
$${\cal F}_J(S)\,, \quad 
\{[{\cal A}_{K_j},\bar{\cal A}_{\bar{K}_j}](F_j)\}_J\,,$$ 
where ${\cal F}_J(S)=\{F_j\}_{j=1,\ldots,J}$ is a partition
of $S$ in $J$ parts, and $[{\cal A}_{K_j},\bar{\cal A}_{\bar{K}_j}](F_j)$
are $J$ irreducible pairs of partitions of the disjoint sets $F_j$, with 
$\cup_{j=1}^J{\cal A}_{K_j}(F_j)={\cal A}_{K}(S)$ and $\cup_{j=1}^J\bar{\cal
  A}_{\bar{K}_j}(F_j)=\bar{\cal A}_{\bar K}(S)$. 
\end{quote}
Here the union of partitions of disjoint sets denotes the union of the 
corresponding families of sets. 
To prove this statement, notice that for any subset $S_1\subseteq S$,
a partition ${\cal A}_{K}(S)$ provides a natural covering of $S_1$,
defined as 
\begin{equation}
  \label{eq:ab-step}
    \Op_{\cal A}[S_1] 
= {\cup}_{k=1}^K\{a_k|a_k\cap S_1\ne \emptyset\}
\,.
\end{equation}
The following properties of $\Op_{\cal A}$ hold: 
\begin{equation}
  \label{eq:ab-step_prop}
  \begin{aligned}
    1.&~~  S_1\subseteq \Op_{\cal A}[S_1]\,; &&&
    2.&~~  \text{if}~S_1\subseteq
    S_2\subseteq S\,,~\text{then}~~\Op_{\cal A}[S_1]\subseteq\Op_{\cal
      A}[S_2]\,;\\  
    3.&~~ \Op_{\cal A}[a_k]=a_k\,;
    &&&     4.&~~  \text{if}~S_1,S_2\subseteq
    S\,,~\text{then}~~\Op_{\cal A}[S_1\cup S_2] =
    \Op_{\cal A}[S_1]\cup\Op_{\cal A}[S_2]\,.
  \end{aligned}
\end{equation}
Fixed points $F=\Op_{\cal A}[F]$ of $\Op_{\cal A}$ coincide with
their covering ({\it self-covering}), 
so they must be of the form 
$F=\cup_{k\in{\cal I}_F}a_k$ for some ${\cal I}_F\subseteq
I_{\cal A}$. Consider next $\Op_{{\cal A}\bar{\cal A}}[F] \equiv
\Op_{\cal A}[\Op_{\bar{\cal A}}[F]]$. It is
straightforward to show that $\Op_{{\cal A}\bar{\cal A}}[F]=F$ if and only
if $F$ is self-covering with respect to both
${\cal A}$ and $\bar{\cal 
  A}$ ({\it biself-covering}), i.e., $F=\cup_{k\in{\cal
    I}_F}a_k=\cup_{\bar{k}\in \bar{\cal 
    I}_F}\bar{a}_{\bar k}$ for some
${\cal I}_F\subseteq I_{\cal A}$  and $\bar{\cal I}_F\subseteq
I_{\bar{\cal A}}$. We call the partitions
$\{a_k\}_{k\in{\cal I}_F}$ and $\{\bar a_{\bar k}\}_{\bar
  k\in\bar{\cal I}_F}$ the {\it induced partitions} of $F$.
If the induced partitions of $F$ form an
irreducible pair, then we say that $F$ is irreducible. By definition,
an irreducible biself-covering set does not contain proper
biself-covering subsets.

The proof now goes as follows. Since, by property 1, 
$\Op_{{\cal A}\bar{\cal A}}^{n}[a_k]\subseteq \Op_{{\cal A}\bar{\cal
    A}}^{n+1}[a_k]$ 
$\forall n\in\mathbb{N}$, and since $S$ is finite, there must be 
$n_k\in\mathbb{N}$
such that $F_{(k)}\equiv\Op_{{\cal A}\bar{\cal A}}^{n_k}[a_k]=
\Op_{{\cal A}\bar{\cal    A}}^{n_k+1}[a_k]=
\Op_{{\cal A}\bar{\cal A}}[F_{(k)}]$, i.e., $F_{(k)}$ is
biself-covering. 
We now show that the induced partitions of $F_{(k)}$
form an irreducible pair, so $F_{(k)}$ is
irreducible. If not, there would be a  
biself-covering proper subset $F'\subset F_{(k)}$, and since also
$F^{\prime\prime}= F_{(k)}\backslash F'$ would be biself-covering, 
we can assume without loss of generality that $a_k\subseteq F'$.
Then, by property 2 in Eq.~\eqref{eq:ab-step_prop}, 
\begin{equation}
  \label{eq:part_abs}
  F_{(k)} = \Op_{{\cal A}\bar{\cal A}}^{n_k}[a_k] \subseteq \Op_{{\cal A}\bar{\cal
      A}}^{n_k}[F'] = F' \subset F_{(k)}\,, 
\end{equation}
which is absurd. 
A similar argument shows that if $a_{k'}\subseteq F_{(k)}$, then 
$F_{(k')}=F_{(k)}$, and analogously $\bar F_{(\bar k)}=F_{(k)}$ if
$\bar a_{\bar k}\subseteq F_{(k)}$, with $\bar F_{(\bar k)}$ generated
from $\bar a_{\bar k}$ as described above. 
Finally, the sets $F_{(k)}$ are
all the irreducible biself-covering subsets of $S$: if $F'$ is an
irreducible biself-covering set, then $\exists a_k\subseteq F'$, and by
property 2 $F_{(k)}\subseteq F'$, which contradicts irreducibility
unless $F_{(k)}= F'$. Obviously $\cup_k F_{(k)}=S$, and so the set 
$\{F_j\}_{j=1,\ldots,J}$ of the $J$ distinct $F_{(k)}$'s provides the
unique partition ${\cal F}_J(S)$ of $S$, such that the induced
partitions of $F_j$, denoted by ${\cal A}_{K_j}(F_j)$ and $\bar{\cal
  A}_{{\bar K}_j}(F_{j})$, form irreducible pairs $[{\cal
  A}_{K_j},\bar{\cal A}_{{\bar  K}_j}](F_{j})$. This completes the 
proof. 

Obviously, to any pair ${\cal F}_J(S)$, $\{[{\cal A}_{K_j},\bar{\cal
  A}_{\bar{K}_j}](F_j)\}_J$, with $[{\cal A}_{K_j},\bar{\cal
  A}_{\bar{K}_j}](F_j)$ 
{\it any} irreducible pair of partitions of $F_j$, corresponds a
unique pair of partitions of $S$, i.e., ${\cal
  A}_{K}(S)\equiv\cup_{j=1}^J{\cal 
  A}_{K_j}(F_j)$ and $\bar{\cal A}_{\bar K}(S)\equiv\cup_{j=1}^J\bar{\cal
  A}_{\bar{K}_j}(F_j)$. 
The sum over pairs of partitions of a set $S$ can
therefore be written equivalently as 
\begin{equation}
  \label{eq:sum_pair_part}
  \sum_{K}\sum_{{\cal A}_K(S)}\sum_{\bar K}\sum_{\bar{\cal A}_{\bar K}(S)}=
  \sum_{J} \sum_{{\cal F}_J(S)} \prod_{F\in{\cal F}_J(S)}
\left(\sum_{K}\sum_{\bar K}\sum_{[{\cal
        A}_{K},\bar{\cal A}_{\bar K}](F)}\right)\,. 
\end{equation}
Consider now the matrices  
\begin{equation}
  \label{eq:matrix}
A^k_{i}=\delta_{k \ell(i)} \,, \qquad
\bar{A}^{\bar k}_{i}=\delta_{\bar k \bar \ell(i)}\,,
\end{equation}
where $\ell(i)$ and $\bar \ell(i)$ associate to each element $i\in S$
the labels of the parts of ${\cal A}_{K}$ and $\bar{\cal A}_{\bar K}$
that contain it. The columns $A^k$ and $\bar{A}^{\bar k}$
are not all
linearly independent, and satisfy exactly $J$ independent relations,
\begin{equation}
  \label{eq:vanish3}
   \sum_{\{k | a_k\in F_j\}} 
A^k_{i} 
= 
\sum_{\{\bar k| \bar a_{\bar k}\in F_j\}} 
\bar A^{\bar k}_{i} 
\,, \quad j=1,\ldots,J\,.
\end{equation}
To see this, define the $J$ linear combinations
\begin{equation}
  \label{eq:vectors2}
Y^{(j)}_{i}(h,\bar h) \equiv
   \sum_{\{k | a_k\in F_j\}} 
A^k_{i} h_k
- \sum_{\{\bar k| \bar a_{\bar k}\in F_j\}} 
  \bar A^{\bar k}_{i}
\bar h_{\bar k}\,,
\end{equation}
which are immediately seen to be linearly independent, as they have no
components in common. There are therefore at most $J$ relations
of linear dependence among columns, of the form
$  Y^{(j)}_{i}(h,\bar h) = 0$, 
which in components read 
\begin{equation}
  \label{eq:vanish4}
h_{\ell(i)} = \bar h_{\bar \ell(i)} ~~\forall i\in
F_j\,.
\end{equation}
We now show that $u_j(i)\equiv h_{\ell (i)} = \bar h_{\bar \ell(i)}$ is
constant over each $F_j$, from which 
Eq.~\eqref{eq:vanish3} follows. 
By definition, $h_{\ell (i)}$ is constant over any $a_k$, 
and similarly $\bar h_{\bar \ell (i)}$ is constant over any
$\bar{a}_{\bar k}$, and so will be $u_j(i)$. 
Suppose now that $u_j(i)$ is constant over a
subset $Q\subseteq F_j$. Then $u_j(i)$ is obviously constant in the
covering of $Q$ provided by ${\cal A}_K$, since $\Op_{\cal A}$
``completes'' the parts already present in $Q$. By the same token,
$u_j(i)$ will also be constant in $\Op_{\bar{\cal A}}[Q]$, and
in $\Op_{{\cal A}\bar{\cal A}}[Q]$.
Since $F_j=\Op_{{\cal A}\bar{\cal A}}^{n_k}[a_k]$
for some $a_k$ and $n_k\in \mathbb{N}$, applying this argument
repeatedly we prove our statement.

As a final comment, consider the matrix obtained by adjoining the 
columns ${\{k| a_{k}\in F_j\}}$ of $A^k_{i}$ 
and ${\{\bar k| \bar a_{\bar k}\in F_j\}} $
of $\bar A^{\bar k}_{i}$. From the result above, its rank is
$K_j+\bar{K}_j-1$. Since the rank has to be smaller than or equal to
the number of rows, i.e., the total number of objects in $F_j$,
$N_{j}$, we have $K_j+\bar{K}_j-1\le N_{j}$.

\section{Exponentiation}
\label{app:expo}

In this Appendix we discuss in some detail the derivation of the 
exponential formula, Eq.~\eqref{eq:expo_final}.
In the previous Appendix we have shown that each pair of partitions 
${\cal A}_K(S_\alpha)$, $\bar{\cal A}_{\bar K}(S_\alpha)$ of $S_\alpha$, with
$K$ and $\bar K$ parts, respectively, 
can be uniquely rewritten as a partition ${\cal F}_J(S_\alpha)$ with
$J$ parts and a set of irreducible pairs of partitions 
$[{\cal A}_{K_j},\bar{\cal A}_{\bar K_j}](F_j)$
of the parts $F_j\in {\cal F}_J(S_\alpha)$.
Using Eq.~\eqref{eq:sum_pair_part}, and dropping temporarily the
dependencies on $b$, $r_{1,2\para}$,  
and $\vec r_{1,2\perp}$ for simplicity, the product 
of two matrix elements $M^{\,(T)}(\Omega_{S_\alpha})$ $\bar
M^{\,(T)}(\Omega_{S_\alpha})$ can be written as 
\begin{equation}
  \label{eq:func_of_part2}
  \begin{aligned}
M^{\,(T)}(\Omega_{S_\alpha})\bar M^{\,(T)}(\Omega_{S_\alpha})
&=\sum_J    \sum_{{\cal F}_J(S_\alpha)}
\prod_{F\in  {\cal F}_J(S_\alpha)} \left\{
\sum_{K}\sum_{\bar K}
\sum_{[{\cal A}_{K},\bar{\cal A}_{\bar K}](F)}
  \tilde M_{[{\cal A}_{K},\bar{\cal A}_{\bar K}](F)}
(\Omega_{F})
\right\} \,,
\end{aligned}
\end{equation}
where
\begin{equation}
  \label{eq:summand_pair}
  \tilde M_{[{\cal A}_{K},\bar{\cal A}_{\bar K}](F)}
(\Omega_{F}) \equiv
\prod_{a\in {\cal A}_{K}(F)} M^{\,(T)\,{\rm conn}}(\Omega_{a})
\prod_{\bar a\in \bar{\cal A}_{\bar K}(F)} 
\bar M^{\,(T)\,{\rm conn}}(\Omega_{\bar a})
\,,
\end{equation}
and we have made us of the fact that, by construction, 
\begin{equation}
  \label{eq:union_mom}
  \bigcup_{a\in {\cal A}_{K}(F)} \Omega_{a} = 
   \bigcup_{\bar a\in \bar{\cal A}_{\bar K}(F)} \Omega_{\bar a} =
   \Omega_{F}\,.
\end{equation}
A partition ${\cal F}_J(S_\alpha)$ of $S_\alpha$ in $J$ parts is fully 
specified by the $J$ strings $\alpha_j\equiv\{N_{s,j}\}$ 
of occupation numbers
$N_{s,j}=N_s(\alpha_j)$ (the number of elements of type $s$ in part
$j$), satisfying 
$\sum_s N_{s,j}\ne 0$ (while $N_{s,j}$ may be zero for some $s,j$) and
$\sum_j N_{s,j}=N_s$, by a reference partition with the given
occupation numbers, and by $n_{\rm sp}^\alpha$ permutations $P_s\in
{\cal S}_{N_s}$, one for each of the $n_{\rm sp}^\alpha$ types
that are present in $S_\alpha$. This  
representation is redundant, with $J!\prod_{s,j} N_{s,j}!$ pairs
$(\{N_{s,j}\},\{P_s\})$ corresponding to the same partition, since the
labeling of the parts is irrelevant and permutations of elements of
the same type 
within a part do not yield a new partition. 
The sum over partitions
${\cal F}_J(S_\alpha)$ can then be written explicitly as
\begin{equation}
  \label{eq:replF}
\sum_{{\cal F}_J(S_\alpha)}
= \f{1}{J!}\sum_{\{\alpha_j\}^\alpha_J} 
\prod_{j=1}^{J} \f{1}{\prod_s N_{s}(\alpha_j)!}  
\sum_{\{P_s\in{\cal
    S}_{N_s}\}}\,, \quad
\sum_{\{\alpha_j\}^\alpha_J} 
\equiv
\sum_{\alpha_1\ne \emptyset}\ldots\sum_{\alpha_J\ne \emptyset}  
\prod_s\left[\delta_{\sum_{j'=1}^J N_{s,j'},N_s} \right]\,.
\end{equation}
Consider now the phase-space integral of
$M^{\,(T)}(\Omega_{S_\alpha})\bar M^{\,(T)}(\Omega_{S_\alpha})$.   
Since the integration measure is factorised, we have
\begin{equation}
  \label{eq:ps_int1}
  \begin{aligned}
    \lella
\prod_{F\in {\cal F}_J(S_{\alpha})} \tilde M_{[{\cal A}_{K},\bar{\cal A}_{\bar K}](F)}
(\Omega_{F})  
\rirra_{\Omega_{S_\alpha}}
= \prod_{F\in {\cal F}_J(S_{\alpha})}
    \lella
\tilde M_{[{\cal A}_{K},\bar{\cal A}_{\bar K}](F)}
(\Omega_{F}) 
\rirra_{\Omega_{F}}\,,
  \end{aligned}
\end{equation}
and so 
\begin{equation}
  \label{eq:ps_int1bis}
  \begin{aligned}
    G_{S_\alpha}^{(T)} & =    \lella
M^{\,(T)}(\Omega_{S_\alpha})\bar M^{\,(T)}(\Omega_{S_\alpha})
\rirra_{\Omega_{S_\alpha}}\\
&=
\sum_J\sum_{{\cal F}_J(S_\alpha)}
\prod_{F\in  {\cal F}_J(S_\alpha)} 
 \left\{ 
\sum_{K}\sum_{\bar K}\sum_{[{\cal A}_{K},\bar{\cal A}_{\bar K}](F)}
\lella \tilde M_{[{\cal A}_{K},\bar{\cal A}_{\bar K}](F)}
(\Omega_{F}) 
\rirra_{\Omega_{F}}
\right\}\,.
  \end{aligned}
\end{equation}
Taking into account that particles of the same type are
indistinguishable, the sum over permutations in Eq.~\eqref{eq:replF}
can be carried out trivially, and after a relabeling of the particles
we get 
\begin{equation}
\label{eq:ps_int2}
  \begin{aligned}
\f{G_{S_\alpha}^{(T)}}{\prod_s N_{s}!}
&=
\sum_J\f{1}{J!}\sum_{\{\alpha_j\}^{\alpha}_J} 
\prod_{j=1}^J
\left\{\f{1}{\prod_s N_s(\alpha_j)! 
}
{\cal Q}_{\alpha_j}^{(T)}
\right\}\,,
  \end{aligned}
\end{equation}
where 
\begin{equation}
  \label{eq:expo_final_app}
  \begin{aligned}
{\cal
  Q}_\alpha^{(T)}
&=
\sum_{K}\sum_{\bar K}\sum_{[{\cal A}_{K},\bar{\cal A}_{\bar K}](S_\alpha)}
\lella
  \tilde M_{[{\cal A}_{K},\bar{\cal A}_{\bar K}](S_\alpha)}
(\Omega_{S_\alpha})
\rirra_{\Omega_{S_\alpha}}\,,
  \end{aligned}
\end{equation}
with the sum being over irreducible pairs of partitions only. 
Summing now over states with different particle content, and using
standard combinatorics results, we finally obtain
\begin{equation}
  \label{eq:ps_int3}
  \begin{aligned}
    {\cal G}^{(T)}(b;r_{1\para},\vec{r}_{1\perp},r_{2\para},\vec{r}_{2\perp})  
&= \exp \left\{ \sum_{\alpha\ne \emptyset}\f{1}{\prod_s N_{s}(\alpha)!}
{\cal Q}_\alpha^{(T)}(b;r_{1\para},\vec{r}_{1\perp},r_{2\para},\vec{r}_{2\perp}) 
\right\} 
\,,
  \end{aligned}
\end{equation}
where we have reinstated the full notation.

\section{Contributions to the potential}
\label{app:largeT}

In this Appendix we compute the contributions 
$V_{[{\cal A}_{K},\bar{\cal A}_{\bar K}](S_\alpha)}$ to the static
dipole-dipole potential, defined 
in Eq.~\eqref{eq:pot_contrib}. Recall that
\begin{equation}
  \label{eq:pot_contrib_app}
  \begin{aligned}
    &-V_{[{\cal A}_{K},\bar{\cal A}_{\bar
        K}](S_\alpha)}(b;r_{1\para},\vec r_{1\perp},r_{2\para},\vec r_{2\perp}) 
  \\ & \equiv  \lim_{T\to\infty}\f{1}{T}\lella 
\prod_{a\in {\cal     A}_K(S_\alpha)}
M^{\,(T)\,{\rm conn}}(\Omega_{a};r_{1\para},\vec r_{1\perp})
\prod_{\bar a\in \bar{\cal A}_{\bar K}(S_\alpha)}
 \bar M^{\,(T)\,{\rm conn}}(\Omega_{\bar{a}};r_{2\para},\vec r_{2\perp})
\rirra_{\Omega_{S_\alpha};\,b}\\
& = \lim_{T\to\infty}\f{1}{T} \int d\Omega_{S_\alpha}\,
e^{-bE(\Omega_{S_\alpha})}
\prod_{a\in {\cal A}_K(S_\alpha)}
\int dt_a e^{iq_a t_{a}}F_T(t_a,P_{a};r_{1\para},\vec r_{1\perp}) \!\!
\\ &\phantom{ \lim_{T\to\infty}\f{1}{T} \int d\Omega_{S_\alpha}\,
e^{-zE(\Omega_{S_\alpha})}\,}
\times 
\prod_{\bar{a}\in \bar{\cal A}_{\bar K}(S_\alpha)}
\int d\bar{t}_{\bar a} e^{-i\bar{q}_{\bar a} \bar{t}_{\bar a}} 
\left[F_T(\bar{t}_{\bar a},P_{\bar a};r_{2\para},-\vec r_{2\perp})\right]^* 
\,,  
  \end{aligned}
\end{equation}
where we have introduced the quantity
\begin{equation}
  \begin{aligned}
& F_T(t_a, P_{a}; r_\para,\vec r_\perp) \equiv
\\
&= {\rm Lim}_{a} \prod_{\is\in a} [-i\pi^{(s)}(\vec
p_{\is},s_{3\is})]
\left[\int d\hat{X}_a\,e^{i {\cal R}_a(\hat X_{a},P_{a})} \,
C_T(t_{a},\hat{X}_{a}; r_\para,\vec r_\perp) \right]_{P_{a\,4}\to
e^{-i\f{\pi}{2}}(-P^0_{a})}\,, 
  \end{aligned}
\end{equation}
where $d X_{a}= dt_a\, d\hat{X}_a$, and 
$P_{a}\cdot X_{a} = q_a t_a + {\cal R}_a(\hat X_{a},P_{a})$, 
with $q_a = \sum_{\is\in a} p_{\is\,1}$, and
$ t_{a} = \f{1}{N_{a}}\sum_{\is\in {a}} x_{\is\,1}$ with $N_{a}$ the
number of elements in $a$. The explicit form of the remainder ${\cal
  R}_a(\hat X_{a},P_{a})$ is not needed. Here we have
dropped the subscript $E$ from Euclidean coordinates and momenta
for simplicity. In terms of $F_T$ we have [see Eq.~\eqref{eq:M_FT}]
\begin{equation}
\label{eq:M_FT_app}
M^{\,(T)\,{\rm conn}}(\Omega_{a};r_{1\para},\vec r_{1\perp})=
  \int dt_a\,e^{iq_a t_a } F_T(t_a,P_{a}; r_{1\para},\vec r_{1\perp})\,.
\end{equation}
As discussed in Subsection \ref{sec:Tdep}, for large $T$
\begin{equation}
  \label{eq:largeT_phys_app}
  \begin{aligned}
&\lim_{T\to\infty}  F_T(T\tau_a,P_{a}; r_{1\para},\vec r_{1\perp})=
\chi\left(\tau_a\right){\cal M}^{\,{\rm conn}}(\Omega_{a};r_{1\para},\vec r_{1\perp})
\,,    
  \end{aligned}
\end{equation}
with
\begin{equation}
  \begin{aligned}
&{\cal M}^{\,{\rm conn}}(\Omega_{a};r_{1\para},\vec r_{1\perp})\equiv\\
& {\rm Lim}_{a} \prod_{\is\in a} [-i\pi^{(s)}(\vec
p_{\is},s_{3\is})]
\left[\int d\hat{X}_a\,e^{i {\cal R}_a(\hat X_{a},P_{a})} \,
C(\hat{X}_{a}; r_{1\para},\vec r_{1\perp}) \right]_{P_{a\,4}\to
  e^{-i\f{\pi}{2}}(-P^0_{a})}\,.
  \end{aligned}
\end{equation}
The other connected matrix element, $\bar M^{\,(T)\,{\rm conn}}$ [see
Eq.~\eqref{eq:clust_M_4}], can be similarly recast as
\begin{equation}
  \begin{aligned}
    \bar M^{\,(T)\,{\rm conn}}(\Omega_{\bar a};r_{2\para},\vec r_{2\perp})
&= \int d\bar t_{\bar a}\,e^{-i\bar q_{\bar a} \bar t_{\bar a} }
\left[ F_T(\bar t_{\bar a},P_{\bar a};
  r_{2\para},-\vec r_{2\perp})\right]^*\,,
  \end{aligned}
\end{equation}
where $
\bar t_{\bar a} = \f{1}{N_{\bar a}}\sum_{\is\in {\bar a}} x_{\is\,1}$
and $\bar q_{\bar a} = \sum_{\is\in  {\bar a}} p_{\is\,1}$.
In full analogy with what was done above, we have
\begin{equation}
  \label{eq:barcalM_app}
  \begin{aligned}
\lim_{T\to\infty}  \left[F_T(T\bar\tau_{\bar a},P_{\bar a};
  r_{2\para},-\vec r_{2\perp})\right]^*&=
\chi\left(\bar\tau_{\bar a}\right)\bar{\cal M}^{\,{\rm
    conn}}(\Omega_{\bar a};r_{2\para},\vec r_{2\perp}) 
\,,    \\
\bar{\cal M}^{\,{\rm conn}}(\Omega_{\bar{a}};r_{2\para},\vec r_{2\perp}) &=
\left[{\cal M}^{\,{\rm 
      conn}}(\Omega_{\bar{a}};r_{2\para},-\vec r_{2\perp})\right]^*
\,.
  \end{aligned}
\end{equation}
To compute $V_{[{\cal A}_{K},\bar{\cal A}_{\bar K}](S_\alpha)}$,
it is convenient to change variables and use $K+\bar K-1\le {\cal
  N}_\alpha$ out of the $K+\bar K$ linear combinations $q_a$ and
$\bar{q}_{\bar a}$, which we denote collectively with $q$ and $\bar
q$, and other $3{\cal N}_\alpha-(K+\bar K-1)$ linearly independent 
combinations of the momenta, which we denote collectively with $\hat
P$. 
That only $K+\bar K-1\le {\cal N}_\alpha$ of the $q_a$
and $\bar{q}_{\bar a}$ are independent follows from the results of Appendix
\ref{app:partitions}. Indeed, in the notation of Appendix
\ref{app:partitions}, 
$q =  A p_{1}$ and $\bar q = \bar A p_{1}$, with $p_{1}$ denoting
collectively all the $p_{i_s\,1}$, 
and the matrix
obtained by adjoining the columns of $A$ and $\bar A$ has rank $K+\bar
K-1$.   
The Jacobian of the change of variables can be chosen to be unity, and
so we can write the phase-space integration measure as
\begin{equation}
  \begin{aligned}
\int d\Omega_{S_\alpha}=
\sum_{\{s_3\}}\int\prod_{\is\in S_\alpha}\f{d^3p_{\is}}{(2\pi)^3}
&= 
\int d\hat P\!
\int{d q}\!
\int {d \bar q}
\,2\pi\delta\left(\sum_a q_a-\sum_{\bar a} \bar q_{\bar a}\right)\,,
  \end{aligned}
\end{equation}
where
\begin{equation}
  \label{eq:intmeasqbarq}
{d q} = \prod_{a\in {\cal A}_K} \f{d q_a}{2\pi}\,,
\qquad
{d \bar q} = \prod_{\bar a\in \bar{\cal A}_{\bar K}} \f{d \bar q_{\bar a}}{2\pi}\,,  
\end{equation}
and $\int d\hat P$ is understood to include also the summation over
spin, which plays no role in the following.
We now set
\begin{equation}
  \label{eq:feffe}
  \begin{aligned}
{\cal F}_T(t,\bar t,&q,\bar q,\hat P;b,
r_{1\para},\vec r_{1\perp},r_{2\para},\vec r_{2\perp}) \\ &=
e^{-bE(\Omega_{S_\alpha})} \prod_{a\in {\cal A}_K} F_T(t_a,P_{a};
r_{1\para},\vec r_{1\perp})
\prod_{\bar a\in \bar{\cal A}_{\bar K}}  \left[F_T(\bar t_{\bar
    a},P_{\bar a}; r_{2\para},-\vec r_{2\perp})\right]^*\,. 
  \end{aligned}
\end{equation}
Dropping the dependence on $b$, and on size and orientation of the
dipoles, ${\cal F}_T$ 
behaves as follows at large $T$, 
\begin{equation}
  \label{eq:largeTcalF}
\lim_{T\to\infty}{\cal F}_T(T\tau,T\bar \tau ,q,\bar q,\hat P) =
{\cal F}(q,\bar q,\hat P) \prod_{a\in {\cal A}_K} \chi(\tau_a)
\prod_{\bar a\in \bar{\cal A}_{\bar K}} \chi(\bar \tau_{\bar a})\,.  
\end{equation}
With this notation, and using the integral representation of the
Dirac delta, we can write
\begin{equation}
  \begin{aligned}
&-V_{[{\cal A}_{K},\bar{\cal A}_{\bar K}](S_\alpha)}
\\ & = \lim_{T\to\infty} \f{1}{T}
\int
d\hat P \!\int 
dq \!\int 
d\bar q \!\int d\omega \!\int dt \!\int d\bar
t \,
e^{i\left[\sum_a q_a (t_{a}-\omega) -
\sum_{\bar a} \bar{q}_{\bar a}  (\bar{t}_{\bar a}-\omega)
\right]}
{\cal F}_T(t,\bar t,q,\bar q,\hat P)
\,.
  \end{aligned}
\end{equation}
Rescaling now $q_a,\bar q_{\bar a}\to q_a/T,\bar q_{\bar a}/T$,
$t_a,\bar t_{\bar a}\to t_a T,\bar t_{\bar a} T$, and $\omega\to
T\omega$, and using the large-$T$ behaviour 
of ${\cal F}_T$, we find
\begin{equation}
\label{eq:QT}
  \begin{aligned}
& - V_{[{\cal A}_{K},\bar{\cal A}_{\bar K}](S_\alpha)}
\\ &=\int
d\hat P \!\int dq\!\int d\bar q\!\int dt\!\int d\bar t\!\int d\omega \,
{\cal F}(0,0,\hat P) \prod_{a\in {\cal A}_K} \chi(t_a)e^{i q_a
  (t_{a} -\omega)}
\prod_{{\bar a}\in \bar{\cal A}_{\bar K}} \chi(\bar t_{\bar a})
e^{-i\bar{q}_{\bar a} (\bar{t}_{\bar a}-\omega)}\\
 &= 
\int
d\hat P \!\int d\omega \,
{\cal F}(0,0,\hat P) \chi(\omega)^{K+\bar K}  
=
\int
d\hat P \,
{\cal F}(0,0,\hat P)
\,.
  \end{aligned}
\end{equation}
Changing integration variables back to the original ones, this
expression can be recast  in the following equivalent, but
physically more clear form, 
\begin{equation}
 \label{eq:delta_fin_app}
  \begin{aligned}
&-V_{[{\cal A}_{K},\bar{\cal A}_{\bar K}](S_\alpha)}
(b;r_{1\para},\vec r_{1\perp},r_{2\para},\vec r_{2\perp})   
\\ &= 
\int
d\hat P\!\int dq \!\int d\bar q
\,{\cal F}(q,\bar q,\hat P)\, 2\pi\delta\left({\textstyle \sum_a} q_a 
\right)
\prod^\circ_{a\in {\cal A}_K(S_\alpha)} 2\pi\delta(q_a)
\prod^\circ_{{\bar a}\in \bar{\cal A}_{\bar K}(S_\alpha)} 2\pi\delta(\bar q_{\bar a})\,
 \\
 &= 
\int d\Omega_{S_\alpha}\,e^{-bE(\Omega_{S_\alpha})}\,
(2\pi)^{K+\bar K-1} \delta_{[{\cal A}_{K},\bar{\cal A}_{\bar K}](S_\alpha)}
(p_1)
  \tilde M_{[{\cal A}_{K},\bar{\cal A}_{\bar
      K}](S_\alpha)}(\Omega_{S_\alpha};r_{1\para},\vec
  r_{1\perp},r_{2\para},\vec r_{2\perp}) 
\,,
  \end{aligned}
\end{equation}
where $\tilde M_{[{\cal A}_{K},\bar{\cal A}_{\bar K}](S_\alpha)}$ has been defined
in Eq.~\eqref{eq:summand_pair}, and
\begin{equation}
  \label{eq:delta_AB}
  \begin{aligned}
&\delta_{[{\cal A}_{K},\bar{\cal A}_{\bar K}](S_\alpha)}
(p_1) \equiv \\
& 
\delta\left({\textstyle\sum}_{\is\in
      S_\alpha}p_{\is\,1}\right)
\prod^\circ_{a\in {\cal A}_K(S_\alpha)}
\delta
\left({\textstyle\sum}_{\is\in
      a}\,p_{\is\,1}\right)
\prod^\circ_{{\bar a}\in \bar{\cal A}_{\bar K}(S_\alpha)}
\delta \left({\textstyle\sum}_{\is\in
      \bar a}\,p_{\is\,1}\right)
\,,
  \end{aligned}
\end{equation}
where the symbol $\circ$ denotes that the product is over all the
parts in the partition but one.

\section{Selection rules for spin-zero particles}
\label{sec:spzero}

In this Appendix we derive the selection rule $\eta_P=\eta_C=1$ for the
Wilson-loop matrix element corresponding to a state with a single,
self-conjugate spin-zero particle. To
this end, we first notice the following transformation 
laws for the Wilson-loop operator 
$\hat\W_E(r_\para,\vec r_\perp; \hat u)$ under charge conjugation, $C$, and
parity, $P$ 
[here we have made explicit also the dependence on the orientation of
the ``long'' side, $u_E=(1,0,0,0)=(\hat u,0)$]:
\begin{equation}
  \label{eq:transflaws}
  \begin{aligned}
   \hat U(C) \hat\W_E(r_\para,\vec r_\perp; \hat u)\hat U(C)^\dag
&= \hat\W_E(r_\para,\vec r_\perp; -\hat u) = 
\hat\W_E(-r_\para,-\vec r_\perp; \hat u) \,, \\
\hat U(P) \hat\W_E(r_\para,\vec r_\perp; \hat u) \hat U(P)^\dag 
&= \hat\W_E(r_\para,-\vec r_\perp; -\hat u) =
\hat\W_E(-r_\para,\vec r_\perp; \hat u)\,.
  \end{aligned}
\end{equation}
Notice that ``time'' is chosen again in the direction of the spatial
separation $\vec b$ between the dipoles. 
Under rotations, $R$, one has in general $\hat U(R)\hat
\W_E[\C]\hat U(R)^\dag=\hat\W_E[R\C]$, with an obvious meaning of the
notation. In particular, for 
$\hat\W_E(r_\para,\vec r_\perp; \hat u)$ and for 
rotations of $\pi$ radians around the axes $\hat r_\perp$ and $\hat u$,
denoted, respectively, by $R_\perp$ and  $R_u$, we have 
\begin{equation}
  \label{eq:transflaws3}
  \begin{aligned}
   \hat U(R_\perp) \hat\W_E(r_\para,\vec r_\perp; \hat u)\hat U(R_\perp)^\dag &=
   \hat\W_E(r_\para,\vec r_\perp; -\hat u) =    
   \hat U(C) \hat\W_E(r_\para,\vec r_\perp; \hat u)\hat U(C)^\dag \,, \\
   \hat U(R_u) \hat\W_E(r_\para,\vec r_\perp; \hat u)\hat U(R_u)^\dag &=
   \hat\W_E(r_\para,-\vec r_\perp; \hat u) = 
   \hat U(P)\hat U(C) \hat\W_E(r_\para,\vec r_\perp; \hat u)\hat
   U(C)^\dag \hat U(P)^\dag\,. 
  \end{aligned}
\end{equation}
For states $|\Omega_{S_\alpha}\ra=|\vec p\,\ra$ containing a single 
spin-zero particle, the relevant
matrix element, 
$M(\vec p;r_{\para},\vec r_{\perp})$, must be of the form
\begin{equation}
  \label{eq:transflaws4}
  M(\vec p;r_{\para},\vec r_{\perp})=
  \delta\left(\vec p\cdot \hat u\right) f(\vec p\cdot \vec
  r_\perp,\vec p^{\,2}; r_\para,\vec r_\perp^{\,2})\,,
\end{equation}
for some function $f$, as a consequence of rotation invariance, and of
translation invariance along $\hat u$ in the limit $T\to\infty$.
For a self-conjugate particle with
parities $\eta_C$ and $\eta_P$, one has moreover, from
Eq.~\eqref{eq:transflaws3},
\begin{equation}
  \label{eq:transflaws5}
  \begin{aligned}
    M(\vec p;r_{\para},\vec r_{\perp})&= \eta_C
    M(R_\perp\vec p;r_{\para},\vec r_{\perp}) = \eta_C
    M(\vec p;r_{\para},\vec r_{\perp})
    \,, \\
    M(\vec p;r_{\para},\vec r_{\perp}) &= \eta_C\eta_P
    M(-R_u\vec p;r_{\para},\vec r_{\perp}) = \eta_C\eta_P
    M(\vec p;r_{\para},\vec r_{\perp})\,,
  \end{aligned}
\end{equation}
where Eq.~\eqref{eq:transflaws4} was also used. The selection rules
then follow immediately. 

One can further exploit Lorentz invariance of the Minkowskian
Wilson-loop matrix elements to prove that for spin-zero particles of
mass $m$ the
Euclidean matrix elements depend only on $r_\para$ and $\vec
r_\perp^{\,2}$ in the limit of vanishing spatial momentum. Indeed, for
one-particle states 
\begin{equation}
  \label{eq:lorentz}
  M_M(\vec p;r_{\para},\vec r_{\perp}) = \delta\left(\vec p\cdot \hat
    u\right)  F_M(p\cdot R_M,R_M^2) = \delta\left(\vec p\cdot \hat
    u\right)  F_M(p^0r_\para - \vec p\cdot\vec r_\perp,r_\para^2-\vec r^{\,2}_\perp)\,,
\end{equation}
and after the Wick rotation $r_\para\to -ir_\para$
\begin{equation}
  \label{eq:lorentz2}
  M(\vec p;r_{\para},\vec r_{\perp}) = \delta\left(\vec p\cdot \hat
    u\right)  F_M(-ip^0r_\para - \vec p\cdot\vec r_\perp,-r_\para^2-\vec
  r^{\,2}_\perp) \mathop \to_{\vec p \to 0}  \delta\left(\vec p\cdot \hat
    u\right)  F_M(-im r_\para,-\vec  r^{\,2})
\,.
\end{equation}
For two-particle states, $M(\vec p_1,\vec p_2;r_{\para},\vec r_{\perp})$,
one similarly has 
\begin{equation}
  \label{eq:lorentz3}
  \begin{aligned}
 M_M(\vec p_1,\vec p_2;r_{\para},\vec r_{\perp}) &= \delta\left((\vec
    p_1+\vec p_2)\cdot \hat 
    u\right)  F_M(p_1\cdot p_2,p_1\cdot R_M,p_2\cdot R_M,p_1\cdot
  u_M,R_M^2)\,,  \\ 
M(\vec p_1,\vec p_2;r_{\para},\vec r_{\perp}) &
\mathop\to_{\vec p\to 0}
\delta\left((\vec
    p_1+\vec p_2)\cdot \hat 
    u\right)  F_M(m^2,-im r_\para,-im r_\para,0,-\vec r^{\,2})\,.
  \end{aligned}
\end{equation}

\end{document}